\documentclass[12pt,preprint]{aastex}

\usepackage{amsmath}
\usepackage{ccaption}
\newcommand{\dosuu}{{}$^\circ$}
\newcommand{\Zsun}{\hbox{\it Z$_\odot$}}
\newcommand{\Msun}{\hbox{\it M$_\odot$}}
\newcommand{\Msunyr}{\hbox{M$_\odot$~yr$^{-1}$}}
\newcommand{\ergs}{\hbox{erg$\,$sec$^{-1}$}}
\newcommand{\Myr}{Myr}
\newcommand{\Gyr}{Gyr}
\newcommand{\kms}{km~sec$^{-1}$}
\newcommand{\HII}{H\,\textsc{ii}}
\newcommand{\NII}{N\,\textsc{ii}}
\newcommand{\OII}{O\,\textsc{ii}}
\newcommand{\OIII}{O\,\textsc{iii}}
\newcommand{\SII}{S\,\textsc{ii}}

\newcommand{\FeII}{Fe\,\textsc{ii}}
\newcommand{\CoII}{Co\,\textsc{ii}}
\newcommand{\FeIII}{Fe\,\textsc{iii}}
\newcommand{\lam}[1]{\ensuremath{\lambda}\,#1}
\newcommand{\lamlam}[1]{\ensuremath{\lambda\lambda}\,#1}

\newcommand{\csalt}{$c_{salt}$}
\newcommand{\dm}{$\Delta\!m_{15}$}
\newcommand{\ha}{\rm{H}$\alpha$}
\newcommand{\hb}{\rm{H}$\beta$}
\newcommand{\logoh}{$\log_{10}$(O/H)+12}
\newcommand{\logten}{$\log_{10}$}
\newcommand{\sia}{\rm{Si}\,\textsc{ii}~{\lam 4130}}
\newcommand{\OHb}{([O\,\textsc{ii}]~{\lam 3727}+[O\,\textsc{iii}]~$\lambda\lambda$4959,5008)/H$\beta$}
\newcommand{\str}{$s(B)$}
\newcommand{\xx}{$x_1$}

\newcommand{\rni}{$^{56}$Ni}
\newcommand{\ebv}{$E(B-V)$}
\usepackage{color}

\shorttitle{Properties of SN Ia Lightcurves and Host Spectra}
\shortauthors{Konishi et al.}

\begin{document}

\title{Dependences of Type Ia Supernovae Lightcurve Parameters on the Host Galaxy Star Formation Rate and Metallicity}

\author{Kohki Konishi\altaffilmark{1,2}, 
David Cinabro\altaffilmark{3},
Peter M. Garnavich\altaffilmark{4},
Yutaka Ihara\altaffilmark{5}
Richard Kessler\altaffilmark{6,7}, 
John Marriner\altaffilmark{8},
Donald P. Schneider\altaffilmark{9},
Mathew Smith\altaffilmark{10},
Harold Spinka\altaffilmark{11},
J. Craig Wheeler\altaffilmark{12}, 
Naoki Yasuda\altaffilmark{2,13}}
\email{kohki@icrr.u-tokyo.ac.jp}

\altaffiltext{1}{Institute for Cosmic Ray Research, 5-1-5 Kashiwa-no-ha, Chiba, Japan}
\altaffiltext{2}{Department of Physics, Graduate School of Science, University of Tokyo, Tokyo 113-0033, Japan}
\altaffiltext{3}{Department of Physics and Astronomy, Wayne State University, Detroit, MI 48202 USA}
\altaffiltext{4}{University of Notre Dame, 225 Nieuwland Science Hall, Notre Dame, IN 46556, USA}
\altaffiltext{5}{Institute of Astronomy, Graduate School of Science, University of Tokyo, 2-21-1 Osawa, Mitaka, Tokyo 181-0015, Japan}
\altaffiltext{6}{Kavli Institute for Cosmological Physics, The University of Chicago, 5640 South Ellis Avenue, Chicago, IL 60637, USA.}
\altaffiltext{7}{Department of Astronomy and Astrophysics, The University of Chicago, 5640 South Ellis Avenue, Chicago, IL 60637, USA.}
\altaffiltext{8}{Center for Particle Astrophysics, Fermi National Accelerator Laboratory, Batavia, IL 60510, USA}
\altaffiltext{9}{Department of Astronomy and Astrophysics, 525 Davey Laboratory, Pennsylvania State University, University Park, PA 16802, USA.}
\altaffiltext{10}{Astrophysics, Cosmology and Gravity Centre (ACGC), Department of Mathematics and Applied Mathematics, University of Cape Town, Rondebosch 7701, Cape Town, South Africa}
\altaffiltext{11}{Argonne National Laboratory, 9700 S. Cass Avenue, Lemont, IL 60437}
\altaffiltext{12}{Astronomy Department, University of Texas, Austin, TX 78712, USA}
\altaffiltext{13}{Institute for the Physics and Mathematics of the Universe, University of Tokyo, Kashiwa 277-8582, Japan}

\begin{abstract}
We present the dependences of the properties of type Ia Supernovae (SNe Ia) 
on their host galaxies by analyzing the multi-band lightcurves of 118 
spectroscopically confirmed SNe Ia observed by the Sloan Digital Sky Survey 
(SDSS) Supernova Survey and the spectra of their host galaxies.  
We derive the equivalent width of the {\ha} emission line, 
star formation rate, and 
gas-phase metallicity from the spectra and compare these with the 
lightcurve widths and colors of SNe Ia.  In addition, we compare 
host properties with the deviation of the observed distance modulus 
corrected for lightcurve parameters from the distance modulus 
determined by the best fit cosmological parameters. This allows us 
to investigate uncorrected systematic effects in the magnitude 
standardization. 
We find that SNe Ia in host galaxies with a higher star formation rate 
have synthesized on average a larger {\rni} mass and show 
wider lightcurves.  The {\rni} mass dependence on metallicity is 
consistent with a prediction of \citet{tim03} based on nucleosynthesis. 
SNe Ia in metal-rich galaxies ({\logoh}$>8.9$) have become 0.13 $\pm$ 0.06
magnitude brighter after corrections for 
their lightcurve widths and colors, which corresponds to up to 
6 \% uncertainty in the luminosity distance.
We investigate whether parameters for standardizing SN Ia maximum magnitude
differ among samples with different host characteristics.
The coefficient of the color term is larger by 0.67 $\pm$ 0.19 for SNe Ia 
in metal-poor hosts than those in metal-rich hosts when no color cuts are imposed. 
\end{abstract}


\section{Introduction}
Type Ia Supernovae (SNe Ia) show diversity in their optical properties. 
The range of B-band peak luminosity is more than a factor of two.  A 
series of lightcurve widths and colors have been demonstrated 
\citep{phi93,phi99,tri98}.  The radioactive element {\rni} is 
explosively synthesized by the nuclear fusion of carbon and oxygen in 
SN Ia progenitors \citep{tru67,col69,hil00}.  The radioactive decay of 
this element is the major source of the SN Ia luminosity.  
For a decade there have been investigations of the link between the 
properties of SNe Ia and the galaxies which host these SNe Ia 
\citep[e.g.][]{ham00,gal05,how09}.

\citet{ham00} used 44 nearby SNe Ia and their hosts to claim that 
bright SNe Ia occur preferentially in young stellar environments by 
examining the trend between the decline rate of luminosity and the 
color (B-V) of their hosts.  They also claimed that bright SNe Ia 
occur in low luminosity hosts by the examination of the luminosity 
decline rate and the host V-band magnitude.  \citet{gal05} showed a 
tentative trend of fainter SNe Ia for metal-rich hosts.  The star 
formation activity and metallicity of SN Ia host galaxies may also 
affect lightcurve properties of SNe Ia. 
These findings, however, have a large uncertainty due to an 
insufficient sample size and should be updated with a larger set of data. 
Recently there has been a focus on the {\rni} mass.  \citet{how09} and 
\citet{nei09} used over 100 pairs of SNe Ia and their host photometry 
to examine the dependence of the synthesized {\rni} mass on stellar 
metallicity. 
If it can be assumed, as has been done by previous researchers, that 
when a host galaxy is metal-rich, a SN Ia progenitor in the galaxy is 
also metal-rich, one can link SN Ia characteristics with their progenitors. 
There is a theoretical prediction based on the nucleosynthesis: the mass 
of {\rni}, a doubly-magic nucleus, becomes smaller and SN luminosity is
lowered for SN Ia progenitors with larger metallicity, because of a larger 
fraction of neutron-rich nuclei $^{22}$Ne \citep{tim03}.  This prediction 
was suggested by an analytic model and supported by detailed simulations 
\citep{tra05,roe06}. 
Observations have been consistent with this predictions \citep{how09,nei09}.

SNe Ia are one of the best cosmological standard candles 
\citep[e.g.][]{rie98,per99}.  Considerable effort has been put into 
the standardization of maximum luminosity.  Lightcurve properties such 
as stretch and color have been used to determine cosmological 
parameters \citep[e.g.][]{kow08,kes09,hic09b}. 
Recent studies \citep{sul10,kel10,lam10b} reported that SNe Ia are 
brighter in massive hosts and hosts with low star formation rate (SFR) 
per stellar mass (specific SFR), after SN Ia maximum brightness have 
been corrected using their lightcurve shape and color.
These results suggest host properties such as the host stellar mass 
can be treated as well as stretch and color to estimate a distance 
modulus \citep{guy10}.  A recent simulation also suggests the 
possibility of systematic dependence of SN brightness on the 
progenitor metallicity \citep{kas09}.

Several studies have been conducted to investigate lightcurve properties 
and their dependences on their host properties for nearby and high-z 
SNe Ia.  However, nearby samples tend to be biased toward luminous hosts, 
while high-z host spectra have yet to been investigated. 
The Sloan Digital Sky Survey (SDSS) -II Supernova Survey \citep{fri08} 
has performed a three year observation and spectroscopically confirmed 512 
SNe Ia in the intermediate redshift range $0.05<z<0.4$. 
Around 20 \% of SN Ia host galaxies were observed spectroscopically
  by the SDSS \citep{yor00}.  With these data, we present relations 
between SN Ia lightcurve and their host gas properties.  The dependences 
of the {\rni} mass and the Hubble residuals are also presented. 
The data are presented in \S \ref{data09d}.  The determination of host 
gas properties, {\rni} masses, and the Hubble residuals are described 
in \S \ref{measure09d}, and the sample for analysis is defined in 
\S \ref{sample}.  We present the results in \S \ref{result09d}, 
related discussion in \S \ref{disc09d} and the conclusions in \S \ref{sum09d}.
The adopted solar abundance is {\logoh} $=8.66$ from \citet{asp05}.
We also adopt $\Omega_M = 0.281$, derived from only the SDSS-II first year 
cosmology sample under a spatially flat cosmological model with a 
constant dark energy equation of state parameter (a sample for the 
spatially flat cosmological model with constant dark energy equation of state 
parameter; in \citet{kes09}). 
The Hubble parameter is set to be $H_0 = 72$ {\kms}Mpc$^{-1}$. 

\section{Data} \label{data09d}
\begin{figure}
 \plotone{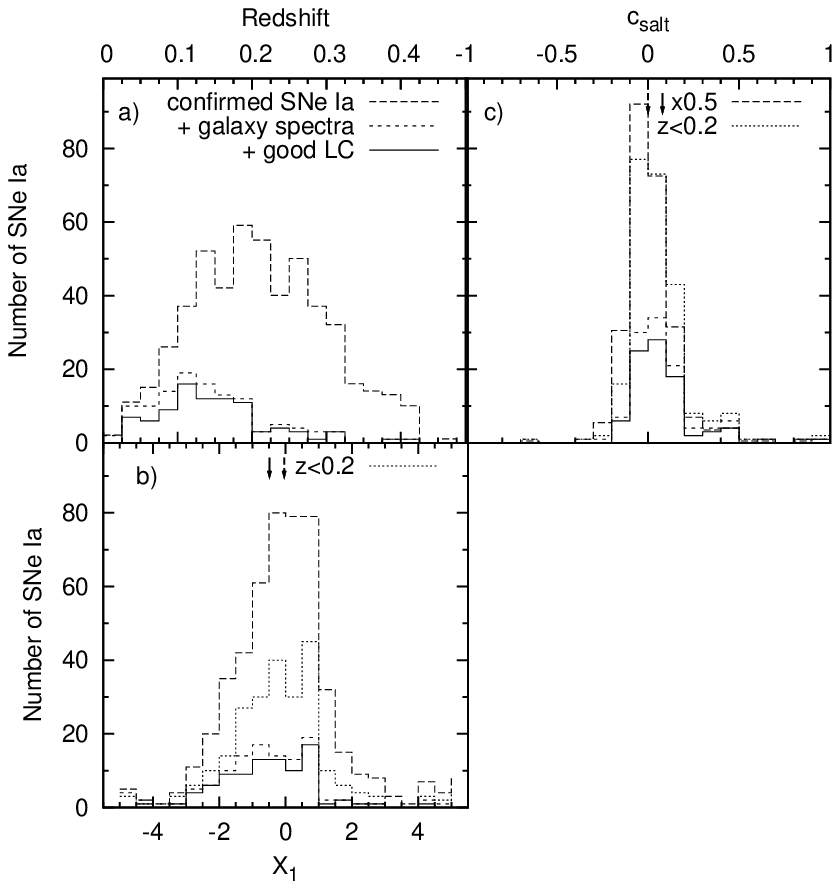}
 \caption{Distributions of (a) redshift, (b) lightcurve width {\xx}, and 
 (c) color {\csalt} distributions for 512 spectroscopically confirmed 
 SNe Ia (bold dashed line), 118 of them with SDSS galaxy spectra (thin 
 dashed lines), and 86 of them that pass all the lightcurve criteria 
 (solid lines; good LC).  The arrows in the {\xx} and {\csalt} 
 distributions are the averages of confirmed SNe Ia and the good LC 
 sample.  The average lightcurve parameters for the good LC sample 
 show a lower {\xx} and a higher {\csalt} relative to the confirmed 
 SNe Ia.  The color histogram for 512 confirmed SNe Ia is multiplied by 
 0.5 for illustrative purposes. 
 Dotted histograms for {\xx} and {\csalt} distributions are confirmed 
 SNe Ia at $z<0.2$.  The {\xx} distributions for SNe Ia at $z<0.2$ and 
 the good LC sample come from the same distribution with a 68\% 
 probability and the {\csalt} distribution with a 98 \% probability. 
\label{lcsum}}
\end{figure}

The SDSS-II Supernova Survey \citep{fri08,sak08} identified 512 spectroscopically confirmed SNe Ia at $0.05<z<0.4$, with lightcurves in five ($ugriz$; \citet{fuk96}) bands from the SDSS 2.5m telescope \citep{gun06} and camera \citep{gun98}.
The survey area is Stripe 82, the 300 deg$^2$ southern equatorial stripe of the SDSS footprint, 20 h to 4 h in right ascension and -1.25~{\dosuu} to +1.25~{\dosuu} in declination. 
Figure \ref{lcsum}a shows the redshift distribution for confirmed 
SNe Ia (dashed line), those with galaxy spectra (dotted line), and those 
plus good lightcurves (solid line, see also \S \ref{gamma}). 
The photon contribution from the galaxy component has been subtracted via the scene modeling photometry method \citep{hol08}.
The sensitivities of the $u$ and $z$ filters are considerably lower than those of the $gri$ bands, so our lightcurve analysis is restricted to the $gri$ bands.

An important aspect of the SDSS-II Supernova Survey is that a larger 
fraction of SN Ia host galaxies were observed spectroscopically than 
SNe Ia discovered by other rolling search surveys. 
Spectroscopic observations were performed with the fiber spectrograph mounted 
on the SDSS 2.5m telescope. 
The fiber aperture was 3 arcsec in diameter and the fiber positions were selected 
to obtain spectra centered on galaxy cores. 
See \citet{sto02} for a description of the SDSS galaxy targeting and algorithm.
The spectral coverage is 3800 to 9200 {\AA} and the wavelength bin is set to 69 {\kms} per pixel in a log-lambda scale; the instrumental resolution is $1850 - 2200$.

In order to identify SN Ia host galaxies, a search was conducted in the SDSS galaxy catalog 
for the closest one in isophotal radius using an exponential profile for the galaxy light. 
Based on comparing redshifts of host galaxies with the redshifts of spectroscopically identified SNe Ia, we estimate that the probability of a SN Ia not being properly matched with its galaxy host at less than a few percent.
As a result of the SN-galaxy matching, we have emission line fluxes for 118 host galaxies. The redshift distribution of the sample is presented in Figure \ref{lcsum}a (thin dashed line).

\begin{figure}
 \plotone{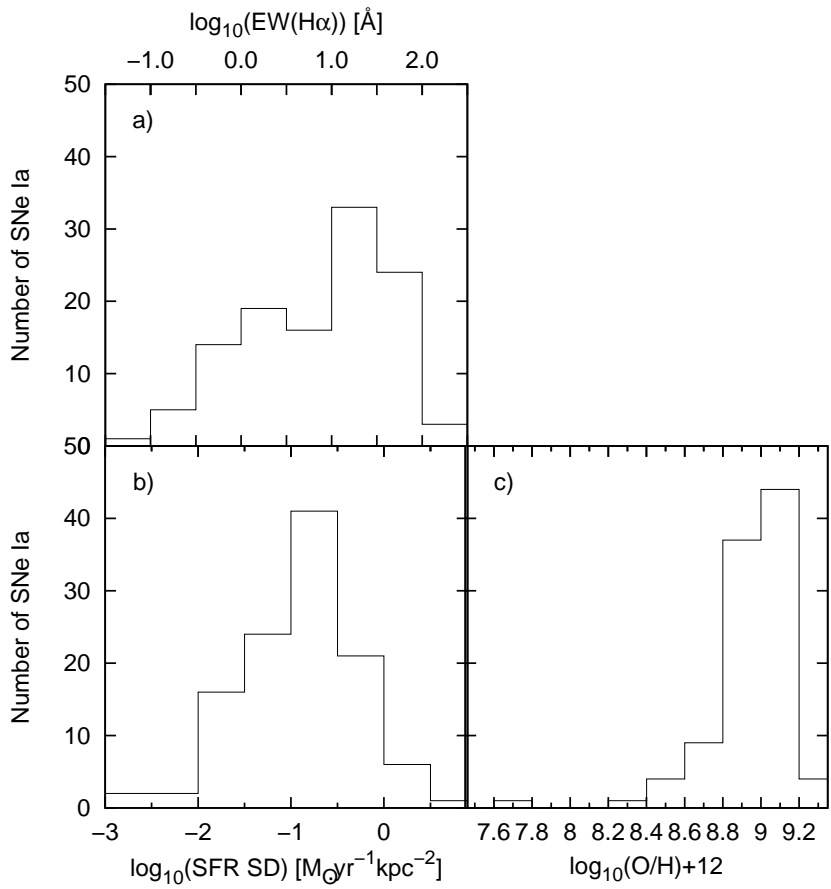}
 \caption{Distributions of (a) EW {\ha}, (b) SFR surface density (SFR SD), 
and (c) metallicity for host galaxies from our measurements. \label{specsum}}
\end{figure}

In order to examine environmental properties on SNe Ia, we used the emission line flux measurements\footnote{\url{http://www.mpa-garching.mpg.de/SDSS/DR7}} by the  MPA/JHU group available to the public. Their spectral sample includes: 
objects brighter than Petrosian $r=17.77$ in the Data Release 7 \citep{aba09}
with (i) {\tt SPECTROTYPE = TARGETTYPE = 'GALAXY'}, a redshift less than 
0.7 and a median signal-to-noise ratio (S/N) per pixel larger than zero 
for a sky-subtracted spectrum.  The sky flux was occasionally 
over-subtracted 1-2 \% and S/N can be below zero (Jarle Brinchmann; 
private comm.), or (ii) {\tt SPECTROTYPE = 'GALAXY'} if the redshift is 
larger than 0.7 and S/N per pixel larger than 2, or (iii) 
{\tt SPECTROTYPE = 'QSO'}.
They measured line fluxes as follows: 
stellar continuum spectra of several different ages and metallicities were generated by a population synthesis code (Charlot \& Bruzual 2010 in prep).
Then a $\chi^2$ fit was performed to construct the best fit continuum for each galaxy spectrum.
After subtracting the best fit continuum from the observed spectra, 
line fluxes were determined by fitting those lines with Gaussians 
simultaneously.
We averaged line fluxes when a galaxy was observed more than once. 
Figure \ref{specsum}a is the histogram of the equivalent width (EW) of the {\ha} emission line (EW {\ha}) 
for our sample. The distribution of EW {\ha} has its peak at the 10-30 {\AA} bin.
These measurements are used to derive star formation rate surface density (SFR SD, whose calculation is described in \S \ref{sfrdef}), and gas-phase metallicity.
Two kinds of possible biases might be included for our sample, one of which arises in the targeting of SNe for spectroscopic confirmation, and the other in the selection for host spectra in SDSS-I.

In Figure \ref{n2ha_o3hb}, we show the characteristics of the interstellar matter (ISM) for SN Ia host galaxies. We plot 77 of 118 SN Ia hosts with
(i) a S/N above two for the {\hb}, [{\OIII}]{\lam 5008}, {\ha} and [{\NII}]{\lam 6585} line, and 
(ii) a redshift greater than 0.04 to avoid the domination of galaxy core components (maximum fraction 20 \%) by fiber aperture effects \citep{kew08}. 
Over 170,000 field galaxy observations were placed in 0.1 bins in {\logten}([{\NII}]{\lam 6585}/{\ha}) and {\logten}([{\OIII}]{\lam 5008}/{\hb}), and the contours connected bins with approximately 100, 2500, 5000, 7500 and 10000 galaxies (black contours from outer to inner). SN Ia host galaxies are shown in red.
This is a diagnostic plot used to separate star forming galaxies and AGN-activity dominated galaxies \citep[e.g.][]{vei87}.
The black dotted curve shows the demarcation between star forming galaxies (left bottom) and AGN-like galaxies (right top) from \citet{kau03}.
Of 77 hosts, 54 galaxies are classified as star-forming galaxies and 23 
as AGN-powered galaxies.
Moreover all the SN Ia host galaxies have the [{\NII}]{\lam 6585}/{\ha} above -1.65, the lower flux ratio of the stellar wind model \citep[discussed below]{hac96}.
We do not correct for reddening in this plot, because the wavelengths of [{\NII}]{\lam 6585} and {\ha} as well as [{\OIII}]{\lam 5008} and {\hb} have only small separations. 

A non negligible fraction of SNe Ia are observed in the outskirts of galaxies \citep{bar07,yas10}. 
The metallicity at SN Ia sites can be estimated by global galaxy luminosity and a metallicity gradient \citep{hen99}.
Although progenitor characteristics for core-collapse SNe may be easily estimated from SN local site information \citep[e.g.][]{boi09}, the estimation is more complicated for SNe Ia.
Several studies have shown a wide range of the SN Ia delay-time, i.e. the time from the birth of a progenitor star to its explosion  \citep[$\lesssim 180${\Myr}; ][]{aub08} to near the cosmic time \citep[$\gtrsim 2.4${\Gyr}; ][]{bra10}. 
For SNe Ia with a long-delay time, local measurements at SN sites are 
probably not representative of the progenitor system, since they have 
the ability to travel significant distances from their star forming regions
due to galaxy random motion and differential rotation.
The mean diameter containing 90 \% of light using Petrosian flux is 6.3 arcsec for our sample, 
which is larger than the fiber aperture. 
Some fibers contain only the light around cores of galaxies, so measured values may not be representative of the global galaxy properties. It is expected that a spatial distribution of these quantities varies among galaxies. We use three galaxy characteristics averaged over the 3 arcsec aperture centered on galaxy cores, metallicity, SFR SD and EW {\ha} for this study. This would be another source of bias.
We include SNe Ia in AGN-like galaxies, following a former study of \citet{gal05}.

\section{Measurements} \label{measure09d}
\subsection{Balmer color excess}
The ratio of {\ha} and {\hb} lines provides an estimate of the color 
excess in a host galaxy assuming a constant intrinsic flux ratio. 

The extinction law $k(\lambda)$ is defined as 
\begin{equation}
 k(\lambda) = \frac{A(\lambda)}{E(B-V)},
\end{equation}
where $A(\lambda)$ is the extinction at the wavelength $\lambda$ in
magnitude and {\ebv} is the color excess in the $B$- relative to the $V$-band.
$R_V = A(V)/E(B-V) = 3.1$ is adopted for our Galaxy.
The extinction is defined as 
\begin{equation}
 A(\lambda) = -2.5 \log_{10}\Bigl( \frac{f(\lambda)}{f_0(\lambda)} \Bigr),
\end{equation}
where $f(\lambda)$ is the observed flux and $f_0(\lambda)$ is the intrinsic
flux without extinction.

The difference in $k(\lambda)$ between $\lambda_1$ and $\lambda_2$ is 
\begin{eqnarray}
 k_1 - k_2 &=& \frac{A_1-A_2}{E(B-V)} \nonumber \\
  &=&-\frac{1.086}{E(B-V)} \ln \frac{f_1/f_2}{(f_1/f_2)_0}.
\end{eqnarray}
In the case of Balmer lines, $k$({\ha}) $-$ $k$({\hb})= $-1.161$ for our Galaxy \citep{cal94}.
There is evidence that this difference is applicable to other galaxies, since the Galactic extinction laws are almost indistinguishable for the Magellanic Clouds within optical wavelengths, $1.0 < 1/\lambda < 4.0$ ${\mu}m^{-1}$ \citep{gor03}.
Although it has been suggested that the ratio of total to selective 
extinction $R_V$ is smaller for SN Ia host galaxies \citep[e.g.][]{nob08}, 
we assume that the extinction law of SN Ia hosts is consistent with that in our Galaxy by following \citet{tak08}.
Intrinsic flux ratios of $f$({\ha})/$f$({\hb}) are presented in \citet{ost89} for optically thin (Case A) and optically thick nebulae (Case B).
Since optically thin nebulae contain only a small amount of gas and therefore are difficult to observe, we use the Case B scenario $f_{int}$({\ha})/$f_{int}$({\hb}) of 2.88. 
The Balmer color excess {\ebv} from Balmer lines is calculated as follows: 
\begin{equation}
 E(B-V) = 0.935 \ln \frac
  {f_{obs}({\rm H\alpha})/f_{obs}({\rm H\beta})}{2.88}.
\end{equation}
The uncertainties are propagated in quadrature.
We assume the dust properties to be like those in our Galaxy and estimate {\ebv} unless the S/N of both line fluxes are below 2 ("N/A" is tagged for non detection and "---" for low S/N cases in Table \ref{spec_sdss}).
The spectra are corrected for the reddening of their Balmer color excess.

\subsection{Star formation rate} \label{sfrdef}

We calibrate the SFR from the {\ha} line, since it is a direct tracer of young stellar populations given by \citet{ken98}
\begin{eqnarray}
  SFR ({\Msunyr}) = 7.9 \times 10^{-42} L({\rm H\alpha}) (\ergs),
\end{eqnarray}
where $L$({\ha}) is the {\ha} luminosity in the galaxy rest frame.

We also estimated the SFR SD by normalizing the SFR by its physical area, 
which is defined as a circle corresponding to the fiber aperture for the 
SDSS spectra.  Figure \ref{specsum}b is the histogram of SFR SD for our 
sample.  
The distribution of SFR SD has its peak in the bin of 0.1 to 0.3 
{\Msunyr}~kpc$^{-2}$.  Table \ref{spec_sdss} lists the results. "N/A" is 
tagged for non-emission galaxies and "---" for low S/N cases.

\subsection{Metallicity}
Oxygen is the most abundant metal in the gas-phase, only weakly
depleted, and exhibits very strong forbidden lines in the optical
wavelength range. 
Ideally, the metallicity is measured from the gas temperature and derived from the flux ratio of the [{\OIII}]{\lam 4364} line to the [{\OIII}]{\lam 5008} line.
However, it can apply only for metal-poor galaxies, since the [{\OIII}]{\lam 4364} line, an auroral line, becomes invisible under metal-rich and cooled environments.
Another method is the usage of strong line ratios.
We use the latter method by following the description of \citet[hereafter KD02]{kew02}, which is reviewed.

KD02 suggests using the flux ratio of $R = $[{\NII}]{\lam 6585}/[{\OII}]{\lam 3727}, where [{\OII}]{\lam 3727} is an abbreviated form of [{\OII}]\lamlam{3726,3729}.
This ratio increases strongly with increasing metallicity for two reasons.
First, since nitrogen is predominantly a secondary stellar nucleosynthesis element while oxygen is a primary one, the [{\NII}] production is roughly proportional to pre-existing seed [{\OII}]. 
Second, the excitation energy of the electron-ion collision is higher for 
[{\OII}]{\lam 3727} than for [{\NII}]{\lam 6585}.
In metal-rich environments, there are fewer thermal electrons with
energy high enough to create the [{\OII}] {\lam 3727} line.
We use Equation 7 of KD02 for the metallicity calibration
\begin{align}
 \log_{10}(O/H)+12 = \log_{10}(1.54020 + 1.26602 R 
	+ 0.167977 R^2)+8.93
  \label{no}. 
\end{align}
If the metallicity is above 0.4{\Zsun}, KD02 suggests that the derived metallicity is taken as a final estimate.
We tagged these galaxies for which this calibration is used as
``N2/O2a'', ``N2/O2d'' or ``N2/O2g'' in the column 7 of Table \ref{spec_sdss};
``a'' is attached for the galaxies with the detection of [{\OII}], {\hb},
[{\OIII}], [{\NII}] and [{\SII}], 
``d'' for those with [{\OII}], {\hb}, [{\OIII}] and [{\NII}],
``g'' for those with only [{\OII}] and [{\NII}].
The R value is corrected based on the reddening correction of the Balmer line ratio.

$R_{23}=$ {\OHb} is also dependent on metallicity, for oxygen is one of the principal nebular coolants. The caveat for this indicator is that the calibration coefficients are model dependent.
Several authors have proposed theoretical calibration schemes, including common ones by \citet{mcg91}, \citet{zar94}, \citet{cha01} and \citet{kob04} (hereafter M91, Z94, C01 and KK04, respectively).
The limitation of the $R_{23}$ ratio is that it gives two values of metallicity (shown {\it upper} and {\it lower} in Equation \ref{m91}). 
Since KD02 claimed that R values are more effectively related to 
metallicity at high metallicities than $R_{23}$, they suggested to 
use $R$ for their high values ($>0.5$ {\Zsun}) and $R_{23}$ for 
small values ($<0.5$ {\Zsun}). 

M91 examined the behavior of $R_{23}$ with metallicity by including the effects of dust and variation in ionization parameter when modelling {\HII} regions.
We use the analytic expression of M91 given in \citet{kob99} to calibrate the metallicity 
\begin{mathletters}
\begin{eqnarray}
 && [\log_{10}(O/H)+12]_{upper} = 12.0 - 2.939 - 0.2 x - 0.237 x^2 
  - 0.305 x^3 - 0.0283 x^4 \nonumber \\
 && - y (0.047 - 0.0221 x-0.102 x^2 - 0.0817 x^3 - 0.00717 x^4) \label{m91} \\
 \nonumber \\
 && [\log_{10}(O/H)+12]_{lower} = 12.0 - 4.944 + 0.767 x + 0.602 x^2
  \nonumber \\
 && - y (0.29 + 0.332 x - 0.331 x^2), 
\end{eqnarray}
\end{mathletters}
where $x = \log_{10}R_{23}$ and 
\begin{equation}
 y = \log_{10}
  \Bigl( \frac{[{\OIII}]{\lamlam 4959,5008}}{[{\OII}]{\lam 3727}} 
  \Bigr).
\end{equation}
Z94 reported that an average of the three calibrations by \citet{edm84},
\citet{mcc85} and \citet{dop86} yields:
\begin{equation}
 \log_{10}(O/H)+12 = 9.265 - 0.33 R_{23} - 0.202 R_{23}^2 
  - 0.207 R_{23}^3 - 0.333 R_{23}^4. \label{z94}
\end{equation}
C01 presents a number of calibrations for various available lines.
Their calibrations are based on a combination of stellar population
synthesis and photoionization codes with a simple model for the dust. 
One of their formulae, also used in KD02, provides the following metallicity relation:
\begin{equation}
 \log_{10}(O/H)+12 = \log_{10} \Bigl[ 5.09 \times 10^{-4} 
  \Bigl( \frac{[{\OII}]/[{\OIII}]}{1.5} \Bigr)^{0.17} 
  \Bigl( \frac{[{\NII}]/[{\SII}]}{0.85} \Bigr)^{1.17} \Bigr]+12. \label{c01}
\end{equation}
KD02 provide a number of calibrations based upon the availability of
particular nebular emission lines. 
KK04 advocate an iterative approach to solve for both quantities
\begin{mathletters}
 \begin{eqnarray}
  && [\log_{10}(O/H)+12]_{upper} = 9.72 - 0.777 x - 0.951 x^2 - 0.072 x^3
   - 0.811 x^4 \nonumber \\
  && - \log_{10}(q) ( 0.0737 - 0.0713 x - 0.141 x^2 - 0.0373 x^3 -
   0.058 x^4) \label{kk04u} \\ 
 \nonumber \\
  && [\log_{10}(O/H)+12]_{lower} = 9.40 + 4.65 x - 3.17 x^2 \nonumber \\
  &&  - \log_{10}(q) (0.272 + 0.547 x - 0.513 x^2) \label{kk04l}.
 \end{eqnarray}
\end{mathletters}
where $q$ is the ionization parameter, determined from
\begin{eqnarray}
 \log_{10}(q) = \{32.81 - 1.153 y^2 
  + [\log_{10}(O/H)+12] (-3.396 - 0.025 y + 0.1444 y^2) \} 
 \\ \nonumber
 \times \{4.603 - 0.3119 y - 0.163 y^2 
  + [\log_{10}(O/H)+12] (-0.48 + 0.0271 y + 0.02037 y^2)\} ^{-1}.
\end{eqnarray}

KD02 compared various calibrations and presented an empirical calibration scheme for metallicity over a wide range. For the galaxies with estimated metallicity below 0.5{\Zsun} from Equation \ref{no}, the average of the M91 formula (Equation \ref{m91}) and Z94 formula should be taken (Equation \ref{z94}). 
If the value is above 0.4{\Zsun}, the derived metallicity is the final estimate.
The symbol ``1b'' is tagged to such galaxies with $>0.4${\Zsun} in the column 7 of Table \ref{spec_sdss}. 
For the galaxies with estimated metallicity below 0.5{\Zsun}, the average of the C01 formula (Equation \ref{c01}) and KK04 formula (Equation \ref{kk04l}) should be taken if they have [{\OII}], {\hb}, [{\OIII}], [{\NII}] and [{\SII}] measurements; these galaxies are tagged with ``1c''.
The Z94 formula (Equation \ref{z94}) should be used for the galaxies with these three emission lines [{\OII}], {\hb} and [{\OIII}] (tagged as ``3f'').
The metallicity for 102 of 118 SN Ia hosts are derived; the remaining 16 hosts either have only {\ha} (and {\hb}) emission line (tagged as ``CaseX") or no lines above S/N$>2$ (tagged as ``N/A").
Figure \ref{specsum}c is the histogram of metallicity for our sample.  The 
distribution of metallicity has its peak in the 9.0 to 9.2 bin.
This might result from the spectroscopic target selection by the SDSS-I 
Legacy survey where only galaxies brighter than $r=17.77$ mag were 
selected.  Bright galaxies have already grown up to be massive and 
metal-rich in the nearby Universe \citep[e.g.][]{tre04}.
Table \ref{spec_sdss} summarizes the spectroscopic properties for the host galaxy spectra. 
The uncertainty in metallicity is calculated by the propagation of uncertainties in line flux measurements. 

\subsection{Ejected {\rni} mass \label{Mni_est}}

{\bf Theoretical prediction} 
The main source of SN luminosity is the decay of the synthesized radioactive {\rni} 
(\citet{tru67},\citet{col69}).
Brighter SNe presumably possess larger {\rni} mass.
\citet{tim03} proposed that less {\rni} mass is created from metal-rich progenitors from their models that conserved charge and mass at the explosion: 
the fusion of $^{12}$C and $^{16}$O triggers the detonation that produces {\rni}, $^{58}$Ni and $^{54}$Fe. 
The electron-to-nucleon ratio and mass fraction after the explosion are
\begin{eqnarray}
 Y_e = \frac{Z(^{56}{\rm Ni}) X(^{56}{\rm Ni})}{A(^{56}{\rm Ni})}
  + \frac{Z(^{58}{\rm Ni}) X(^{58}{\rm Ni})}{A(^{58}{\rm Ni})} \\
 X(^{56}{\rm Ni})+X(^{58}{\rm Ni}) = 1.
\end{eqnarray}
Here $Z(i)$ and $A(i)$ are the number of protons and nucleons (protons
plus neutrons) of the element $i$, respectively, and $X(i)$ is the mass
fraction. 
The relation between the {\rni} mass and the electron-to-neutron ratio is thus 
\begin{equation}
 M(^{56}{\rm Ni}) = 0.6 X(^{56}{\rm Ni}) = 0.6 (58 Y_e - 28) \label{M_ye},
\end{equation}
where a typical {\rni} mass for the $Y_e = 0.5$ progenitor is set to be 0.6 {\Msun}. 
$X(^{54}{\rm Fe})$ was set to be zero for simplification.  Inclusion of 
this element makes the slope of $Y_e$ vs. $M(^{56}{\rm Ni})$ shallower 
by a factor of $(58-56.8)/58 \sim 2$ \%.

{\bf Estimate of $Y_e$} 
In order to derive the {\rni} mass, a straight-forward method is to measure the abundance of SN Ia progenitors. However, this measurement is difficult, since we observe the results of element synthesis in SNe Ia. The best current effort is to use the  metallicity of their hosts. With the definition of the electron-to-nucleon ratio $Y_e$ and the formula of the mass fraction for CO white dwarfs $\sum X(i)=1$, $Y_e$ is expressed as follows \citep{how09}: 
\begin{align}
 Y_e &= \frac{6}{12}X(^{12}{\rm C}) + \frac{8}{16}X(^{16}{\rm O})
  + \frac{10}{22}X(^{22}{\rm Ne}) + \frac{26}{56}X(^{56}{\rm Fe}) \nonumber \\
 &= \frac{1}{2} 
  - \{ X({\rm H}) \Bigl( \frac{{\rm Fe}}{{\rm H}} \Bigr) 
  \Bigl( 3+ \Bigl( \frac{{\rm C}}{{\rm Fe}} \Bigr) \Bigr)
  + X({\rm H}) \Bigl( \frac{{\rm O}}{{\rm H}} \Bigr) 
  \Bigl(2+ \Bigl( \frac{{\rm N}}{{\rm O}} \Bigr) \Bigr) \} \label{metal_ye}. 
\end{align}
We assume a constant (C/Fe) in our metallicity range \citep{whe89}  
and set X(H) and (C/Fe) to be the solar values of 0.7392 and 8.7 \citep{asp05}.
From observations of nearby galaxies or stars within our Galaxy, (N/O) and (Fe/H) 
increase with (O/H) \citep[e.g.][]{pag81,whe89}.  We take the dependence of (O/H) 
on (N/O) from \citet{vil93}: (N/O) = 0.0316 + 126 (O/H) and
on (Fe/H) from \citet{ram07}:
\begin{eqnarray}
\Bigl( \frac{{\rm Fe}}{{\rm H}} \Bigr) = 10^{-a/(1+b)}
\frac{({\rm Fe}/{\rm H})_{\odot}}{({\rm O}/{\rm H})_{\odot}^{1/(1+b)}} 
\Bigl( \frac{{\rm O}}{{\rm H}} \Bigr)^{1/(1+b)},  \label{feh}
\end{eqnarray}
where $a=0.096$ and $b=-0.327$ for the thin disk.  Solar oxygen and 
iron abundances are derived from \citet{asp05}: 
$\log_{10}({\rm O}/{\rm H})_{\odot} = -3.34$ and 
$\log_{10}({\rm Fe}/{\rm H})_{\odot} = -4.55$.
From Equations \ref{metal_ye} and \ref{feh}, the {\rni} mass is represented by a function of (O/H) with four coefficients. It decreases with increasing metallicity.

Prior to the explosion, electron capture by $^{12}$C burning (simmering) can reduce the free electron abundance and therefore reduce the amount of synthesized {\rni}. The effect of electron capture on the variation of {\rni} mass may be small $\lesssim 5$ \% \citep{cham08}. We ignore this simmering effect. 
Although the far UV flux varies with Fe abundance, we 
can neglect this impact on the {\rni} mass, since the flux would only be 
$\sim 3 \times 10^{-3}$ times larger than the optical flux \citep{sau08}.

{\bf Estimate of {\rni} mass} 
We now describe the method to obtain the {\rni} mass from SN Ia lightcurves.
Since the radioactive decay of {\rni} powers the SN Ia luminosity 
(\citet{tru67,col69}) mainly for the photospheric phase, the maximum 
bolometric luminosity $L_{bol}$ is comparable to the radioactive luminosity. 
The {\rni} mass is well described by
\begin{eqnarray}
M(^{56}{\rm Ni}) = \frac{L_{bol}}{\gamma \dot {S}(t_R)},
\end{eqnarray}
where $\dot {S}(t_R)$ is the radioactive luminosity per solar mass of
{\rni} and $\gamma$ is the ratio of bolometric to radioactive luminosity
\citep{arn82}.

Multi-band lightcurves are used to estimate the bolometric luminosity.
We first derive the lightcurve parameters using the SALT2 lightcurve 
fitting code \citep{guy07}.
The SALT2 code employs a two-dimensional spectral surface $F(p,\lambda)$ in time and wavelength constructed by the average temporal evolution of the spectral energy distribution for SNe~Ia ($M_0$) and its deviation ($M_1$). 
\begin{equation}
 F(p, \lambda) = x_0 \times [M_0(p,\lambda)+x_1 M_1(p,\lambda)] 
  \times \exp[c_{salt} CL(\lambda)], \label{fsalt2}
\end{equation}
where $p$ is the rest-frame days from the date of peak luminosity, $x_0$ is the normalization, and {\xx} is the coefficient corresponding to the lightcurve width. 
$CL(\lambda)$ is the average color correction law and {\csalt} is its 
coefficient, which is sensitive to both intrinsic color diversity and 
the host-dust reddening. 
The SALT2 fit returns lightcurve parameters ($x_0$, {\xx}, {\csalt}) for each SN Ia. 

\citet{wan09} presented an extensive dataset of a normal SN Ia (SN 2005cf) from UV to near infrared wavelengths. 
Since the UV flux variation is not fully understood 
\citep{hoe98,len00,sau08}, we assume a negligible variation 
of UV flux among SNe Ia and adopt 0.30 for the fraction of missing flux 
outside the optical window from 2900 to 7000 {\AA} (their Figure 24).
The maximum bolometric flux can thus be estimated using lightcurve 
parameters, spectral surfaces, and the luminosity distance $d_L$,
\begin{equation}
 L_{bol} = \frac{4 \pi d_L^2}{1-0.30} 
  \int_{2900}^{7000} F(0, \lambda) \, d\lambda.
\end{equation}

The radioactive luminosity per solar mass of {\rni} can be estimated 
using $e$-folding decay times for {\rni} $\rightarrow ^{56}$Co and
$^{56}$Co $\rightarrow ^{56}$Fe of 8.8 and 111 days, 
and mean energy release per decay of 1.71 and 3.76 MeV, 
\begin{align}
 \dot {S} &= 6.31 \times 10^{43} e^{-t_R/8.8} \nonumber \\
  & + 1.43 \times 10^{43} e^{-t_R/111} \mbox{{\ergs}{\Msun}$^{-1}$} \label{Sdot},
\end{align}
where $t_R$ is the time from the explosion to maximum B-band brightness (the rise time). The rise time is described by a 'stretch' parameter {\str} which determines broadening or narrowing of an average template \citep{per97,guy05}. Following an average stretch-corrected rise time of $19.5 \pm 0.2$ days \citep{rie99,ald00,gol01,con06}, we set $t_R/s(B) = 19.5$ \citep{how09}. {\str} is derived from the width {\xx} and the polynomial calibration given in \citet{guy07} \footnote{The {\str} parameter is produced directly by \citet{per97} or \citet{guy05}. We believe that the SALT2 code represents SN Ia characteristics more realistically due to a larger training dataset.}.
We have not tried to incorporate the variation in SN rise time reported by \citet{hay10} or use their more precise rise time of $17.38 \pm 0.17$.  Our results are not sensitive to the exact value used, and we prefer to maintain a simple estimate of the {\rni} mass based on the SALT2 light curve parameterization.
The uncertainty in $\dot {S}$ is derived from Equation \ref{Sdot},
\begin{equation}
 \sigma_{\dot{S}}/\sigma_{S} = 
	\left( 1.40 \times 10^2 e^{-2.22 s(B)} + 2.51 \times e^{-0.18 s(B)} \right) \times 10^{42}.
\end{equation}

The quantity $\gamma$ is a correction factor between bolometric to radioactive luminosity.
The peak luminosity is equal to the instantaneous rate of energy deposition by the {\rni} decay assuming constant opacity with time \citep{arn82}.
Since the opacity decreases with the temperature, thermal energy stored in opaque regions is released and adds to the luminosity at later phases i.e. $\gamma > 1$. 
There will be an intrinsic varience in a radial distribution of elements in SNe Ia and this could change the intrinsic variance of $\gamma$ \footnote{In case of SNe Ia with {\rni} distributed toward the outer layer, photons deposited from the outer layer can easily escape and $\gamma$ for such SNe Ia might be small.}. $\gamma$ is thought to be roughly 10 \% \citep{bra95}. We reflect the uncertainty of the intrinsic variance by assigning this ratio to be $1.2 \pm 0.1$ \citep{bra95,how06,how09}.

Following the work of \citet{how09}, the uncertainty in the {\rni} mass is derived by propagating uncertainties in the bolometric luminosity, the radioactive luminosity and the quantity $\gamma$, 
\begin{equation}
 \sigma_{^{56}{\rm Ni}} = \sqrt{ 
	\left( \frac{1}{\gamma \dot {S}} \right)^2 \sigma^2_{L_{bol}} 
	+ \left( \frac{L_{bol}}{\gamma \dot {S}^2} \right)^2 \sigma^2_{\dot {S}} 
	+ \left( \frac{L_{bol}}{\gamma \dot {S}^2} \right)^2 \sigma^2_{\gamma} 
	}, 
\end{equation}
where $\sigma_{\gamma}$ is set to 0.1.

\subsection{Hubble residual}

The Hubble residual ($HR$) is defined as the difference in these two distance moduli:
\begin{eqnarray}
 \mu^{corr}_B 
  &=& \{m^*_B(x_0) + \alpha x_1 - \beta c_{salt}\} - \bar{M}  \label{mcorr} \\
 \mu^{best fit}_B 
  &=& 5\log_{10} \Bigl(\frac{d_L(z, \Omega_M, \Omega_\Lambda)}{10 pc} \Bigr) \\
 HR &=& \mu^{corr}_B - \mu^{best fit}_B,
\end{eqnarray}
where $M$ is the average absolute magnitude of SNe Ia and  $m^*_B$ is the observed peak magnitude.
The $HR$ would be zero for a perfect standard candle, but in practice has an intrinsic scatter of $\sim 0.15$ mag. 
Various efforts have been made to reduce this scatter and randomize it at all redshifts \citep[e.g.][]{phi93,rie96,guy07,jha07,kes09}.

We derive standardization parameters for luminosity $\alpha$, $\beta$ and $M$ so that the $\chi^2 = \sum \bigl(\frac{HR^2}{(\delta \mu_B)^2+\sigma_{int}^2} \bigr)$ is minimized for the parameters. The error in distance modulus $\delta \mu_B$ is calculated by the error propagation of the covariance matrix. The intrinsic dispersion $\sigma_{int}$ is set to be 0.14 mag \citep{lam10b} and is added in quadrature to the error $\delta \mu_B$ to achieve a reduced $\chi^2$ close to one \citep{lam10a}. We use the three samples: SNe Ia with (i) host EW {\ha}, (ii) host SFR SD, and (iii) metallicity.

\section{Sample selection} \label{sample}
We investigate the link between SN Ia lightcurve properties 
and their host properties.  Since our lightcurves have been obtained by a 
period-determined survey, SNe Ia which were discovered near the beginning 
or the end of the period are incomplete.  The following criteria for lightcurves 
were set for the analysis to examine SNe Ia whose lightcurves can be reconstructed 
accurately by the SALT2 fitter:
\begin{enumerate}
 \item at least one data point with $p<-4$,
 \item at least one data point with $p>+4$, 
 \item at least five data points with $-20<p<+60$,
 \item lightcurve parameters with $|x_1|<5.0$ and/or $|c_{salt}|<2.0$, 
\end{enumerate}
where $p$ is the rest-frame phase in days.
30 of a total of 118 SNe Ia do not meet at least one of the criteria above, 23 of which do not meet the first three criteria, and five SNe Ia (SN 12897, 13610, 16644, 18835, 20420) do not meet the fourth criteria. The fourth criterion excludes an additional two SNe Ia which are known to have unusual lightcurves: SN2005hk \citep{phi07} and SN2007qd \citep{mcc10}.

Figure \ref{lcsum} shows the distributions of redshift (panel a), lightcurve widths (panel b) and colors (panel c) of SNe Ia.
The bold dashed histograms are the distributions for 512 spectroscopically 
confirmed SNe Ia.  The thin dashed histograms are those for the 118 with 
host galaxy spectra.  The solid histograms are for the 86 passing the 
lightcurve criteria in addition to having host spectra (the good LC sample). 
Arrows are the average width and color for confirmed SNe Ia 
(bold dashed; $-0.04 \pm 0.07$ and $0.00 \pm 0.01$) and the good LC sample 
(bold solid; $-0.49 \pm 0.15$ and $0.08 \pm 0.02$).  The average values for 
the good LC sample are lower in {\xx} and higher in {\csalt}.  This may be 
because the confirmed sample contains intrinsically bright SNe Ia at high
redshifts: the {\xx} and {\csalt} distributions for the good LC sample and 
the confirmed SNe Ia in a similar redshift range ($z<0.2$; the dotted 
histogram in the panel b) come from the same distribution with a 
68\% and 98 \% probability, respectively. 
The best fit Gaussian to the color histogram of the good LC sample
is $\propto \exp(-((c_{salt}-c_0)^2/2 \sigma_c^2))$, where $c_0 = 0.030$ and 
$\sigma_c = 0.098$.  This Gaussian is used to separate the 
dust-extinguished SNe Ia in \S \ref{gamma}.

In order to the examine dependences of SN lightcurve 
properties on their hosts over a wide range of host properties, we use 
hosts with $EW(H\alpha)/\delta EW(H\alpha)>1$ and 
$f(H\alpha)/\delta f(H\alpha)>1$ for the EW {\ha} and SFR SD sample. 
We divide each sample into sets of equal and sufficient numbers of 
SNe Ia from the highest value of EW {\ha}, SFR SD, or metallicity to 
examine average trends.  The mean, the error on the mean, and the 
deviation (when necessary) for each set are shown as red points in 
\S \ref{gamma} and \S \ref{galcosmo}.  The average value at the 
left-most point results from the remaining SNe Ia.

\section{Results} \label{result09d}
\subsection{The {\rni} mass}
Figure \ref{Mnihist} is the histogram of {\rni} masses for our sample. 
The {\rni} mass ranges from around 1.0 {\Msun} to all the way down close to
zero.  The least {\rni} mass was 0.036 {\Msun} for SN12979.  Note that, 
for SNe Ia with the {\rni} mass less than 0.2 {\Msun}, all the SNe Ia show 
large {\csalt} ($\gtrsim 0.3$). It is likely that the {\rni} mass 
were underestimated due to the dust extinction of their host galaxies. 
The only exception is a SN Ia with the {\rni} mass of 0.06 {\Msun}.
The low value is attributed to a small lightcurve width ({\xx}$\sim 4.0$).
The bolometric flux correction was assumed to be constant for each SN Ia.
This correction is based on only one SN Ia spectrum (\S \ref{Mni_est}).  The 
wide {\rni} mass range, however, can not be explained by varying the 
correction factor.  The range of the correction should be 0.76 (for 
SNe Ia with {\rni} mass of 0.2 {\Msun}) or even negative (-0.16 for 
those of 1.0 {\Msun}) to match to a SN Ia with the {\rni} mass of 0.6 
{\Msun}. 
This scatter has a trend with decline rate \citep{phi93} and ($B-V$) color 
at maximum date \citep{tri98}. 
The left part of Figure \ref{metal_phi} shows the color-corrected 
magnitude $m_B-\beta c-\mu_{best}$ against the width {\xx}.  The right 
part of Figure \ref{metal_phi} shows the width-corrected magnitude 
$m_B+\alpha x_1-\mu_{best}$ against the color {\csalt}.
SNe Ia with the {\rni} mass less than 0.3 {\Msun} or more than 0.8 
{\Msun} are marked in large solid circles.
The linear lines in Figure \ref{metal_phi} are derived by minimizing 
$\chi^2 = \sum \bigl(\frac{HR^2}{(\delta \mu_B)^2+\sigma_{int}^2} \bigr)$.
 SNe Ia with small or large {\rni} masses follow the overall trend.
\citet{kas07} explained these empirical relations as a temperature 
variation: in a cool system, the recombination of {\FeIII} to {\FeII} 
and the development of numerous {\FeII}/{\CoII} absorption lines become 
noticeable at earlier phases in the $B$ band wavelength range.  This 
results in a fast decline of SN Ia brightness.
The number fraction of our sample is the highest in the {\rni} mass 
range from 0.40 {\Msun} to 0.65 {\Msun}. 

\citet{str06} compared two methods for deriving the {\rni} mass.
One method is to obtain bolometric luminosity at maximum brightness 
with a constant $\gamma$. The other is to model Fe features in nebular 
phase spectra. 
All the masses except two out of a total of 17 are consistent within 20 \%.
Since nebular spectra of our SNe Ia can not be observed by current 
instruments, we have adopted the former method described in 
\S \ref{Mni_est}.
This approach yielded a {\rni} mass range from 0.1 {\Msun} for a subluminous 
SN Ia (SN 1991bg) to 1.0 {\Msun}, a comparable mass range to our sample.

\subsection{Environmental effects on lightcurve properties \label{gamma}}

We start with an investigation of relations among SN Ia 
lightcurves and host gas properties: width {\xx} and color {\csalt} for 
SNe Ia, and EW {\ha}, SFR SD, and metallicity for their hosts.
Figure \ref{host_x1} shows dependences of the lightcurve width {\xx} on 
host gas properties: (a) EW {\ha} for 74 SNe Ia, (b) SFR SD for 74 SNe Ia  
and (c) metallicity for 67 SNe Ia.  The metallicity is represented as 
{\logoh} (lower horizontal axis) or [Fe/H] (upper horizontal axis) by 
Equation \ref{feh}.  Note that [Fe/H] is negative for thin disk stars with 
[O/H]$=0$ \citep{ram07}.  The vertical dotted line indicates the value 
of solar metallicity.  SN17332 with the lightcurve properties of ({\xx}, 
{\csalt})$=$(-0.53, 0.11) is eliminated from the following metallicity 
dependence plots because its host metallicity is extremely low {\logoh}$=7.77$. 
The maximum {\xx} appears to be independent of the host properties. 
However, from the data in Figures \ref{host_x1}a and 
\ref{host_x1}b it appears that the lower value decreases from -1.0 for SNe Ia in 
hosts with a large EW {\ha} of around $10^{1.8}${\AA}) or 
with high star formation rate (SFR SD value of $10^0$ {\Msunyr}~kpc$^{-2}$), 
to -2.5 for those in hosts with a low EW {\ha} of around $10^{-1}$ {\AA} or 
with low star formation rate (SFR SD value of $10^{-1.5}$ {\Msunyr}~kpc$^{-2}$).
Moreover, the dispersion of {\xx} for hosts with the low EW {\ha} of the sample 
is 1.1, which is comparable to that for hosts with a high EW {\ha}.
The Kolmogorov-Smirnov (KS) test gives the probabilities that SNe Ia in hosts 
with low and high EW {\ha} come from the same population to be $<1$ \%.
The same result is obtained for the SFR surface density.
The lower value of {\xx} appears to increase from -2.5 for SNe Ia 
in metal-rich ({\logoh} value of $\sim$9.3) hosts to -1.5 for those 
in metal-poor ({\logoh} value of 8.6) hosts (Figure \ref{host_x1}c), 
being the KS probability of 81 \%.

Similarly, Figure \ref{host_c} shows dependences of the 
color {\csalt} on host gas properties: (a) EW {\ha}, (b) SFR SD, and 
(c) metallicity.  The {\csalt} range is essentially constant with 
respect to the EW {\ha} of their hosts (Figure \ref{host_c}a).  If the 
parameter {\csalt} were completely explained by host-dust reddening, 
it would show a correlation with SFR SD.  Figure \ref{host_c}b shows 
wider {\csalt} values (-0.2 to 0.2) for SNe Ia in hosts with modest 
SFR ({\logten}(SFR SD) of -1.5 to 0.5 {\Msunyr}~kpc$^{-2}$) than those 
in high/low star forming hosts.  Some SNe Ia with large {\csalt} 
($\gtrsim$0.2) occur in the hosts with low star forming (SFR SD below 
$10^{-1.5}$ {\Msunyr}kpc$^{-2}$) or high metallicity above 9.0.  
From Figure \ref{host_c}c, it appears that the range of {\csalt} is 
wider (-0.2 to 0.2) for SNe Ia in metal-rich ({\logoh}$>9.0$) hosts than 
for those in metal-poor ({\logoh}$<8.6$) hosts, 0.0 to 0.1, but a 
KS test shows that the two distributions are compatible at the current 
level of statistical accuracy.

The {\csalt} parameter measures SN reddening relative to the nominal SALT2 templates.  
SN reddening can be caused by host galaxy dust extinction, intrinsic variations in the 
SN explosion, its immediate environment or some mixture of them.  
A flat {\csalt} distribution irrespective of the SN radial position \citep{yas10}
and a correlation of the pseudo equivalent width of the ``{\sia}" feature of a SN Ia 
spectrum with {\csalt} \citep{nor10b} support that the intrinsic variation of the SN 
explosion is introduced in {\csalt}.  
We exclude SNe Ia with {\csalt}$>$0.3, larger than 3 $\sigma$ deviation from the averaged
color in our sample (Figure \ref{lcsum}c), since those SNe are most likely reddened 
primarily by host galaxy extinction.

Since the {\rni} mass synthesized in the SN explosion 
is estimated from their lightcurves, it is expected to have some 
dependence on their host properties.  Moreover a metallicity 
dependence can be compared with a theoretical prediction. 
Figure \ref{host_Mni} shows the dependences on (a) EW {\ha}, (b) 
SFR SD, and (c) metallicity. 
The upper values of the {\rni} mass do not change with respect to these 
host properties.   However, from the data in Figures \ref{host_Mni}a 
and \ref{host_Mni}b, it appears that the lower value decreases from 0.4 
{\Msun} for SNe Ia in hosts with a large EW {\ha} and with high star 
formation rate, to 0.2 {\Msun} for those in hosts with a low EW {\ha} 
or with low star formation rate.  The lower value of the 
{\rni} mass appears to increase from 0.25 {\Msun} for SNe Ia in metal-rich hosts 
to 0.4 {\Msun} for those in metal-poor hosts (Figure \ref{host_Mni}c).  
The KS test gives the probabilities that SNe Ia in hosts 
with low and high star formation rate (EW {\ha}, metallicity) 
come from the same population to be 12 \% (38 \%, 84 \%).  This indicates 
that the amount of the {\rni} mass is the most sensitive to the young star 
fraction of their hosts.
Five averaged points show that the {\rni} mass is constant below {\logoh} 
$<8.9$ and that it decreases towards high-metal hosts.  
The average {\rni} mass in the highest metallicity bin is $\sim$0.11 {\Msun} 
smaller than those in the two lowest metallicity bins (but only at $\sim1.6 
\sigma$). A similar trend appears when we divide the dataset into five 
equally-spaced bins (0.16 dex) or when we make six representative points 
ithin the metallicity range of 8.6 to 9.4.
This finding is consistent with the prediction of \citet{tim03} that the 
{\rni} mass decreases by 0.15 {\Msun} for the metallicity range in our 
sample (Equations \ref{M_ye}-\ref{feh}; a blue curve in Figure 
\ref{host_Mni}c).  %Although the result is not statistically significant, 

\subsection{Environmental effects on supernova cosmology \label{galcosmo}}
We derive standardization parameters for maximum luminosity ($\alpha$, $\beta$ 
and $M$) and the Hubble residuals, a linear combination of them with lightcurve 
parameters ({\xx} and {\csalt}) for these samples: EW {\ha}, SFR SD, and metallicity.
Luminosity standardization parameters are derived by minimizing $\chi^2$ 
with the Hubble constant and cosmological parameters fixed.  For this 
analysis, red SNe Ia with {\csalt}$>0.3$ are treated the same as the rest.
The datasets for EW {\ha}, SFR SD, and metallicity consist of 81, 83, and 
72 SNe Ia. 

Table \ref{alpbet} summarizes the best fit luminosity standardization 
parameters and the $HR$ root mean square(rms)s for these datasets.  Hubble 
residuals are derived using these parameters.  Several Hubble residuals 
of a SN Ia are derived if the SN Ia is contained in more than one sample. 
Their dependences on host gas properties are shown in Figure \ref{host_hr} 
for the samples of (a) EW {\ha}, (b) SFR SD, and (c) metallicity.  Five 
averaged Hubble residuals in each panel of Figures \ref{host_hr}a and 
\ref{host_hr}b are consistent with zero ($\sim 1\sigma$).  
We do not observe the average maximum brightness corrected 
for lightcurve shapes and colors to depend on EW {\ha} and SFR SD. 
The Hubble residuals in the two lowest metallicity bins are 0.13 mag 
fainter than those in the three highest metallicity bins (1.8$\sigma$ 
significance).  Even though the significance is marginal, this corresponds 
to up to 6 \% uncertainty in luminosity distance.  

Further, we split each sample by their host characteristics and derived 
luminosity standardization parameters.  Results are again summarized in 
Table \ref{alpbet}.
Figure \ref{alpbetf} shows the significance levels in the differences of 
correction coefficients $\alpha$ (rectangles), $\beta$ (circles) and $M$ 
(triangles) between low/high EW {\ha}, SFR SD and metallicity. Left (right) 
side points of each entry are the values for the sample with (without) the 
color cut. A significance is defined as the difference of a coefficient for 
low/high host properties divided by the root of an error sum of squares. 
Because of the many tests performed (Table \ref{alpbet}), the significances 
on these results are less than what would be naively computed from the number 
of standard deviations. 
(i) SNe Ia in hosts with low EW {\ha} have a marginally larger $\alpha$, 
a marginally smaller $\beta$ ($1.7 \sigma$), and larger negative $M$ 
($3.6 \sigma$) than those in hosts with high EW {\ha}.
(ii) SNe Ia in hosts with low SFR SD have a marginally larger $\alpha$ 
($1.6 \sigma$), a comparable $\beta$, and a marginally smaller negative 
$M$ than those in hosts with high SFR SD.
(iii) SNe Ia in metal-poor hosts ({\logoh} $<9.0$) have a comparable 
$\alpha$ ($<1 \sigma$) to those in metal-rich hosts but have larger 
$\beta$ ($3.5 \sigma$) and marginally smaller negative $M$ 
($1.8 \sigma$) than those in metal-rich hosts. 
(iv) $R_V \sim \beta -1$ is smaller for SN Ia hosts, compared with the 
canonical value of $R_V = 3.1$ for our Galaxy.
(v) The $HR$ rms for SNe Ia in high EW {\ha}, high SFR SD and metal-rich 
hosts were smaller.
These $HR$ rms were comparable to 0.178 mag for the SDSS-II only sample \citep{kes09}.
When these parameters were calculated for the sample without red SNe Ia ({\csalt}$<0.3$), 
the difference in $\beta$ disappears below the $1\sigma$ level, while other parameters 
are barely changed. 

\section{Discussion} \label{disc09d}
The relations between lightcurve characteristics of SNe Ia and their host gas 
properties have been investigated. EW {\ha}, SFR SD and metallicity has been 
used as host gas properties.
For dependences of host galaxies on SN Ia properties, the averaged lightcurve 
width was narrower for SNe Ia occurring in hosts with lower EW {\ha}.  
This suggests that hosts with a lower EW {\ha} give birth to 
SNe Ia with narrower widths of lightcurves.  The lightcurve width was also 
narrower for SNe Ia in lower star forming rate hosts. 
Moreover, the {\rni} mass was observed to have the best sensitivity to the 
EW {\ha} of their hosts. Although this mass shows a wide scatter from $\sim0.2$
to 1.0 {\Msun}, the average trend between the {\rni} mass and metallicity was 
consistent with a theoretical prediction based on the nucleosynthesis \citep{tim03}. 
For the standardization of maximum luminosity, 
the $HR$ averages were constant with respect to EW {\ha} and SFR SD, indicating that 
the dependences of maximum luminosity on them were removed by lightcurve corrections. 
The coefficient $\beta$ was large for SNe Ia in hosts with large EW {\ha} and low metallicity.

{\bf Effects of host galaxies on SN Ia properties: }
Dependences of SN Ia lightcurve properties on host characteristics can 
be compared with a similar study using nearby SNe Ia and their hosts 
\citep{gal05}.  Observing entire spectra of nearby galaxies, they 
measured emission line fluxes and derived EW {\ha}, SFR by the {\ha} 
emission, and metallicity by the ratios of strong lines.  Their result 
shows that the deviation of {\dm} for SNe Ia in hosts with low EW {\ha}
($<18$ {\AA}) is more than twice larger than that for SNe Ia in hosts 
with high EW {\ha}.  
The differences between our findings and theirs probably 
results from sample selections.  Around 80 \% of SN Ia hosts in their 
sample are very bright nearby galaxies categorized before 1980 in the New 
General Catalogue, the Index Catalogue or the Uppsala General Catalogue 
of Galaxies.  %It has not been shown that characteristics of 

Recently the {\rni} mass was correlated with host 
metallicity \citep{how09,nei09} derived from stellar masses and 
empirical calibrations \citep{tre04,lia06}. 
\citet{how09} used over 100 pairs of SNe Ia and their host photometry in 
the high-z universe ($z \sim 0.4$) to report that the average of the 
{\rni} mass becomes smaller for metal-rich hosts.  \citet{nei09} used a 
similar size of SNe Ia and hosts in the local universe (regression 
velocity of $cz\lesssim 30,000${\kms}) and reported that although the data 
were statistically consistent with no trend, they were also consistent 
with the \citet{how09} and the \citet{tim03} model. 
Their method of metallicity estimation is different from ours, since 
emission lines are formed by interstellar gases while photometry is primarily 
contributed by the stellar component.
It can be said that our study has revealed the dependence of the {\rni} mass 
on host {\it gas} metallicity.

{\bf Luminosity standardization: }
\citet{gal05} found insignificant correlations between $HR$ and 
{\ha}-emission related properties of the {\ha} EW and the SFR SD, while 
they showed a slight trend of a negative $HR$ for SNe Ia in metal-rich 
hosts.  However, they cautioned that the trend was less than a 
$2 \sigma$ detection.  Our findings are consistent with theirs. We do 
not make comparisons with \citet{gal08}, which ruled out a 
no-correlation at the 98 \% significance level, because they analyzed 
different type hosts (passive) by a different metallicity estimation 
method (absorption lines).

Recent studies \citep{sul10,kel10,lam10b} reported evidence at between 
2 and $3\sigma$ that SNe Ia are brighter in massive hosts with low 
star formation per stellar mass (specific SFR), after their SN Ia 
maximum brightnesses have been standardized by using their lightcurve 
shapes and colors.  If massive hosts are metal-rich, their results 
agree with brighter SNe Ia in metal-rich hosts after lightcurve 
corrections (Figure \ref{host_hr}c). 

We found a hint that SNe Ia have smaller {\rni} mass on average in metal-rich 
hosts (Figure \ref{host_Mni}c). On the other hand, we found a marginally larger  
negative $M$ (Table \ref{alpbet}) and $HR$ (Figure \ref{host_hr}c) for such SNe Ia. 
Since the {\rni} mass is the main source of 
the SN luminosity \citep{arn82}, smaller {\rni} masses would result in a smaller  
negative $M$.  There is also evidence of smaller negative $M$ for SNe Ia in 
hosts with a high EW {\ha} (Table \ref{alpbet}).  These findings agree with a former study 
\citep{sul10} that SNe Ia in massive hosts becomes 0.08 mag ($\approx 4.0 \sigma$) 
brighter after lightcurve corrections if massive hosts are metal-rich and have 
a low {\ha}.  \citet{kas09} claimed from 
simulations that metal-rich progenitors cause SNe Ia both to be fainter and to 
have narrower lightcurves. They also showed that SNe Ia in metal-rich progenitors 
are brighter for a fixed decline rate of luminosity. 
Thus, our findings are qualitatively consistent with these studies. 

There is evidence that $\beta$ is larger for SNe Ia in metal-poor hosts 
or hosts with a large EW {\ha}.  This suggests that the size of the 
color correction $\beta${\csalt} is larger for SNe Ia with the same 
color in hosts with metal-poor/large EW {\ha} than those with 
metal-rich/low EW {\ha} (Table \ref{alpbet}). 
Similar results have been shown in \citet{lam10b} that $\beta$ is larger for 
SNe Ia in star forming hosts than passive hosts. An uncertainty of 0.16 in the 
$\beta$ estimation for our 36 metal-poor galaxies is comparable to that of 0.16 
for their 40 passive hosts. 
\citet{sul10} derived SFR and stellar mass to present a larger $\beta$ for SNe 
Ia in the hosts with higher specific SFR and with lower stellar mass. 
If these galaxies are metal-poor and show large {\ha} EWs, our results agree with 
theirs.  Since metal-poor and star forming galaxies are dustier than metal-rich 
and passive galaxies, these findings might imply that large dust extinctions in 
hosts increase $\beta$.

As can be seen in Table \ref{alpbet}, there might be evidence for a 
different $\alpha$ between low and high SFR SD hosts at between 
1 and $2\sigma$.
If the effective $\alpha$ does depend on star formation rate, one would expect 
it also to be a function of redshift due to the increase in cosmic star formation 
with redshift \citep{lil96,mad98,cow99,flo99,haa00,wil02,hop04,hop06} would 
introduce a change in $\alpha$. 

\begin{figure}
 \plotone{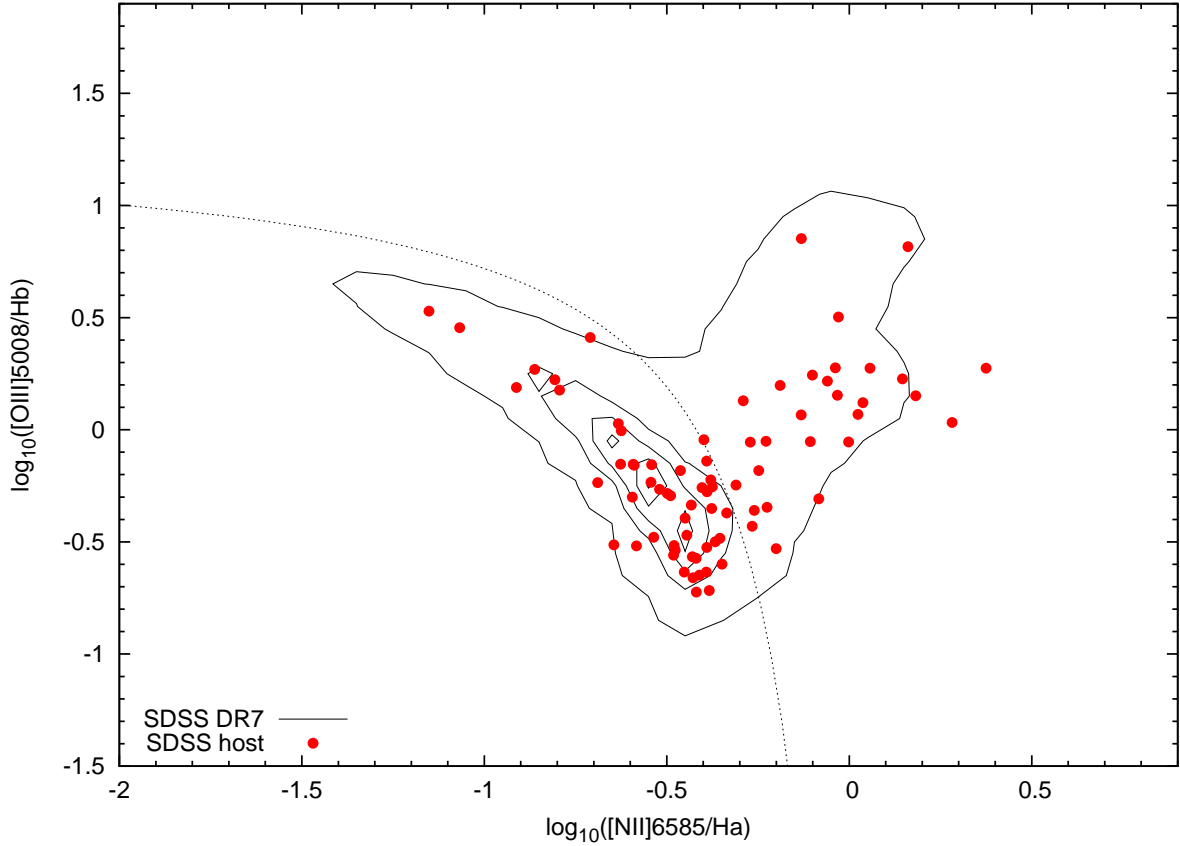}
 \caption{[{\NII}]{\lam 6585}/{\ha} vs. [{\OIII}]{\lam 5008}/{\hb} flux ratios.
The distributions of over 170,000 galaxies for the MPA/JHA sample are shown as 
the black contour lines (100, 2500, 5000, 7500, 10000 galaxies from outer to inner), 
whereas SN Ia host galaxies are in red.
The black curve shows the demarcation between star forming galaxies (left bottom) 
and AGN-like galaxies (right top) from \citet{kau03}.
\label{n2ha_o3hb}}
\end{figure}

\begin{figure}
 \plotone{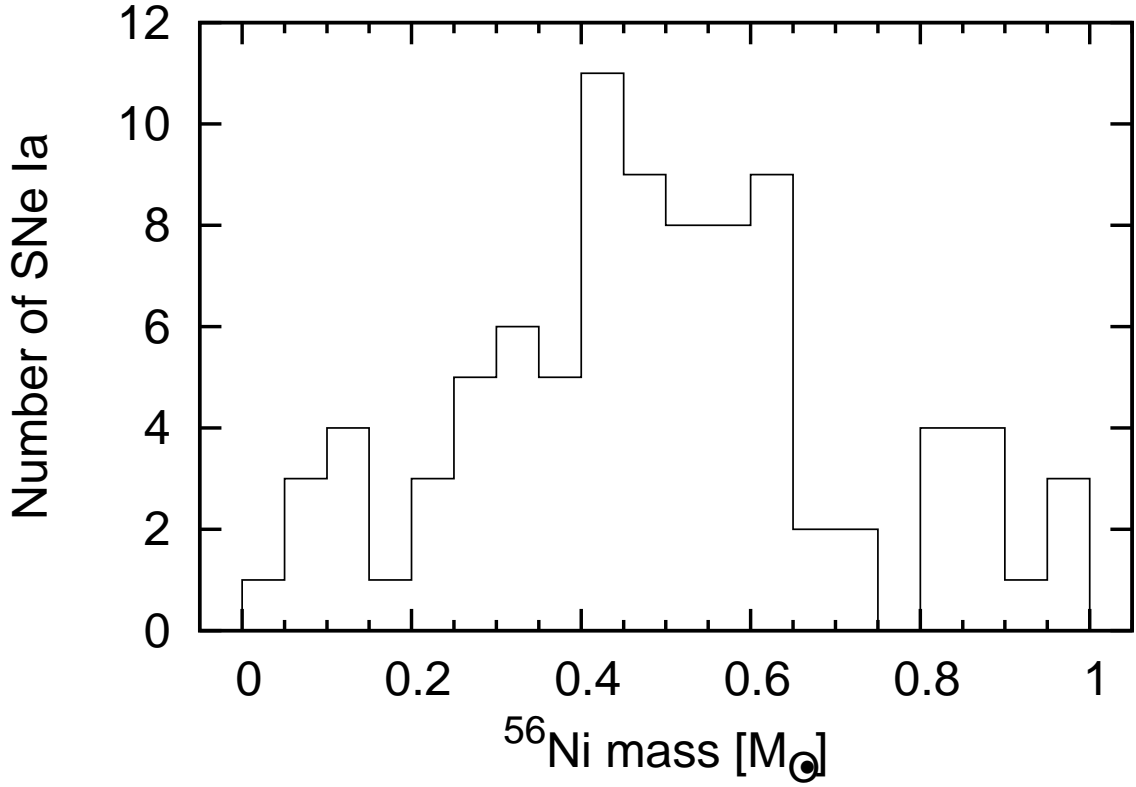}
 \caption{The distribution of {\rni} masses for our sample for the good 
 LC sample.
 \label{Mnihist}}
\end{figure}

\begin{figure}
 \plotone{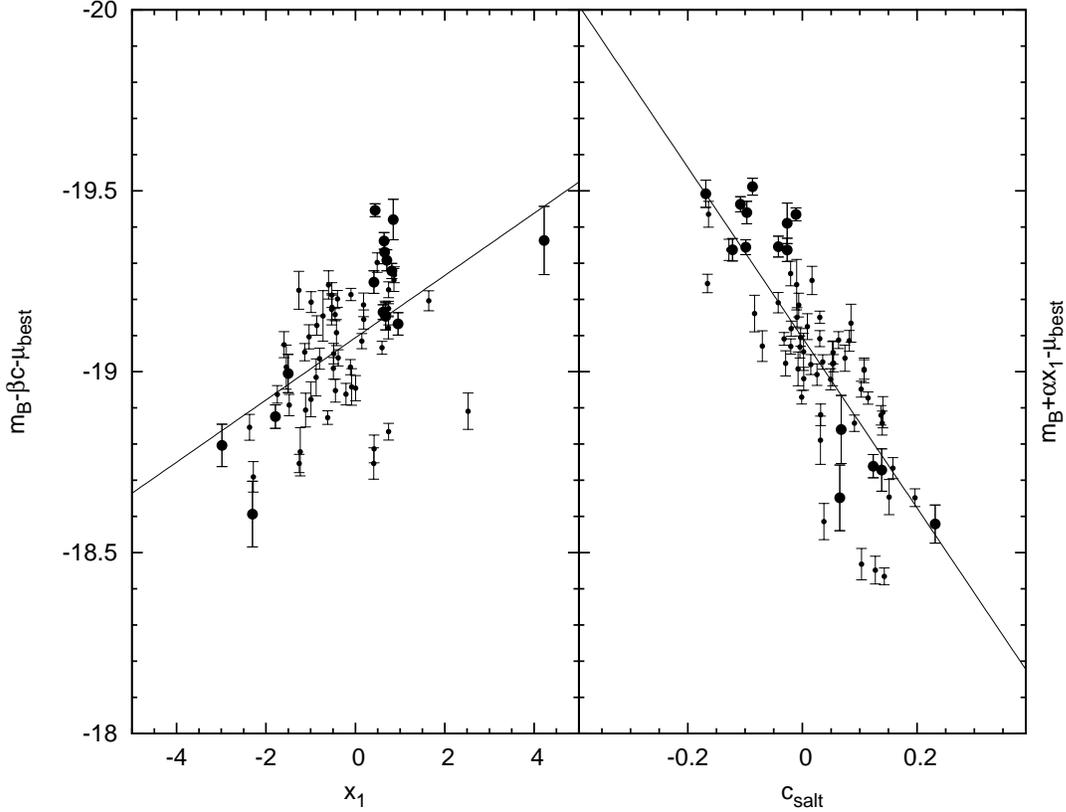}
 \caption{The magnitude-width and magnitude-color relations. 
Left: the color-corrected magnitude $m_B-\beta c-\mu_{best}$ is plotted against the width {\xx}.
Right: the width-corrected magnitude  $m_B+\alpha x_1-\mu_{best}$ is plotted against the color {\csalt}.
SNe Ia with the {\rni} mass $<$0.3 {\Msun} or $>$0.8 {\Msun} are marked with circles, while the others with dots.
The lines in the figures have the values of $\alpha$, $\beta$ and $M$ which minimize $\chi^2 = \sum \bigl(\frac{HR^2}{(\delta \mu_B)^2+\sigma_{int}^2} \bigr)$.
\label{metal_phi}}
\end{figure}

\begin{figure}
 \plotone{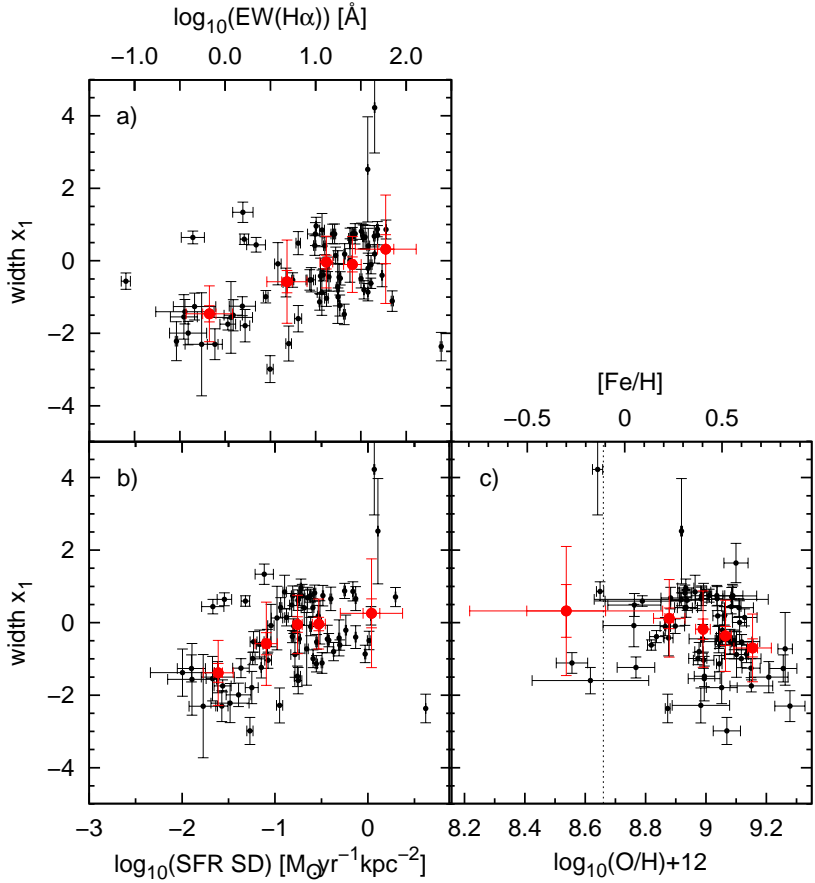}
 \caption{The dependences of the SN Ia lightcurve width {\xx} on host 
 gas properties: (a) {\ha} EW, (b) SFR SD, and (c) metallicity. 
 The metallicity is represented as {\logoh} (lower scale) 
 or [Fe/H] (upper scale).  Note that [Fe/H] is negative for thin disk 
 stars with [O/H]$=0$.  The vertical dotted line indicates the value of 
 solar metallicity.
 Our sample is plotted as black dots. The red points are the mean widths, 
 the error on the mean, and the deviation for each set of 15 SNe Ia from 
 the highest value of host EW {\ha}, SFR SD, and metallicity.  One SNe Ia 
 with the lowest metallicity host of {\logoh}$<8.2$ is eliminated from 
 the figure for illustrative purposes.
\label{host_x1}}
\end{figure}

\begin{figure}
  \plotone{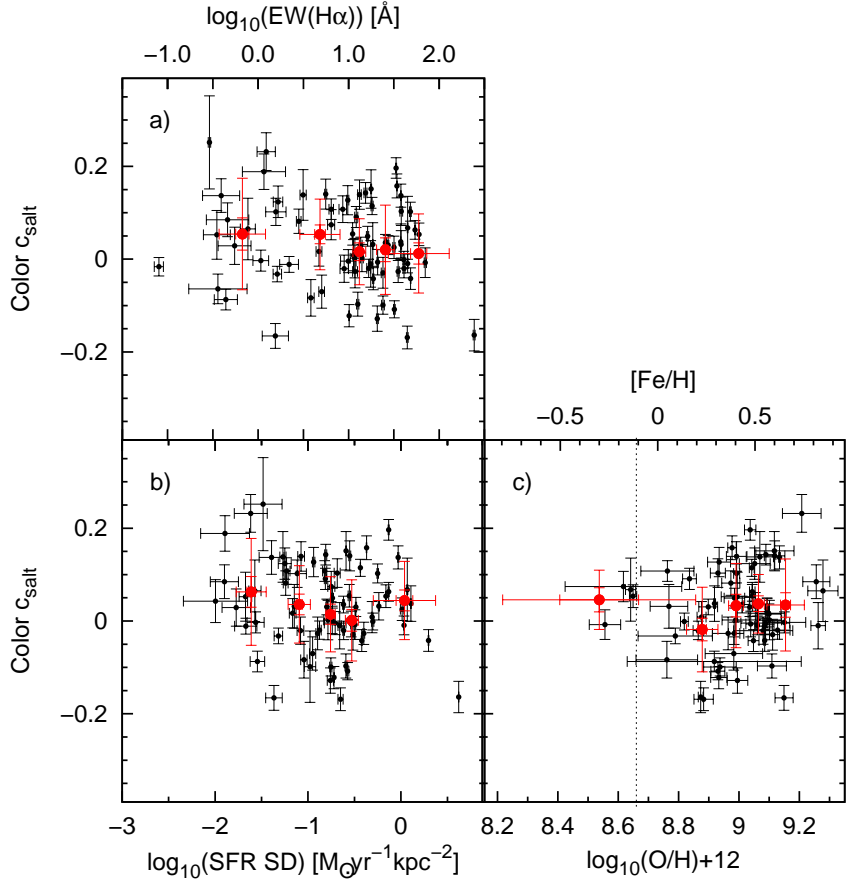}
 \caption{The dependences of the SN Ia color {\csalt} on host gas properties: 
(a) {\ha} EW, (b) SFR SD and (c) metallicity. The symbols are the same as Figure \ref{host_x1}.
\label{host_c}}
\end{figure}

\begin{figure}
  \plotone{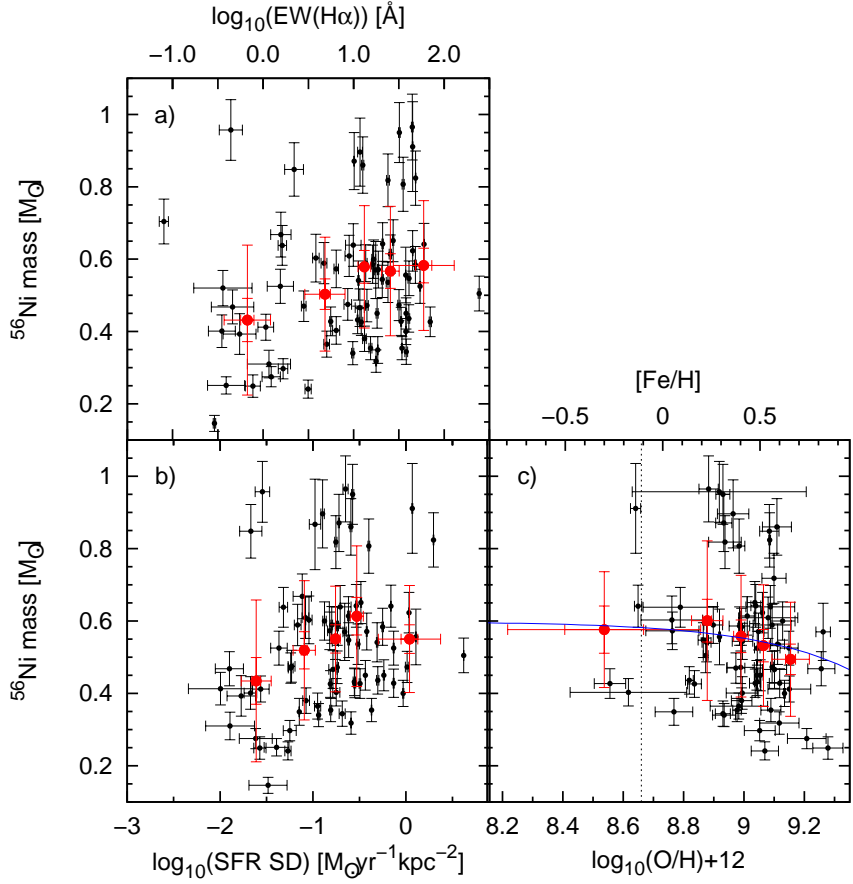}
 \caption{The dependences of the {\rni} mass on host gas properties: 
(a) EW {\ha}, (b) SFR SD and (c) metallicity. The symbols are the same 
as Figure \ref{host_x1}.
The blue curve is the theoretical prediction of \citet{tim03}. 
\label{host_Mni}}
\end{figure}

\begin{figure}
  \plotone{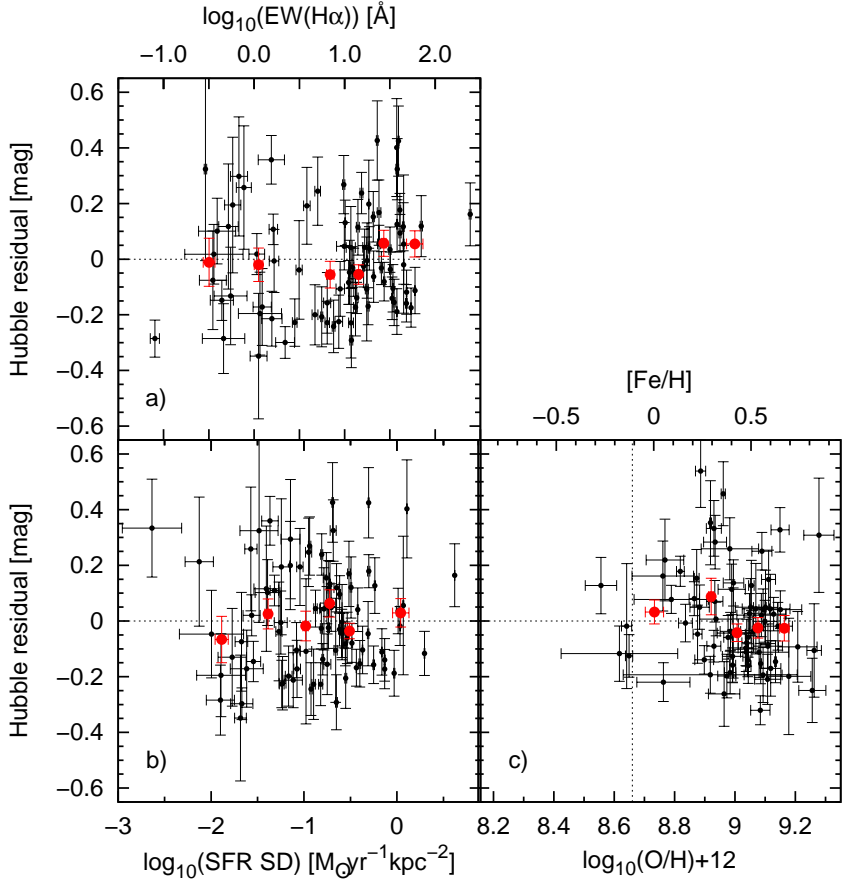}
 \caption{The dependences of the Hubble residuals on host gas properties: (a) {\ha} EW, (b) SFR SD and (c) metallicity. The symbols are the same as Figure \ref{host_x1}.
 Five averaged Hubble residuals in each panel of (a) and (b) are consistent with zero ($\sim 1\sigma$). The Hubble residuals in the two lowest metallicity bins are 0.13 mag fainter than those in the three highest metallicity bins ($\sim1.8\sigma$ significance; panel c).
\label{host_hr}}
\end{figure}

\begin{figure}
  \plotone{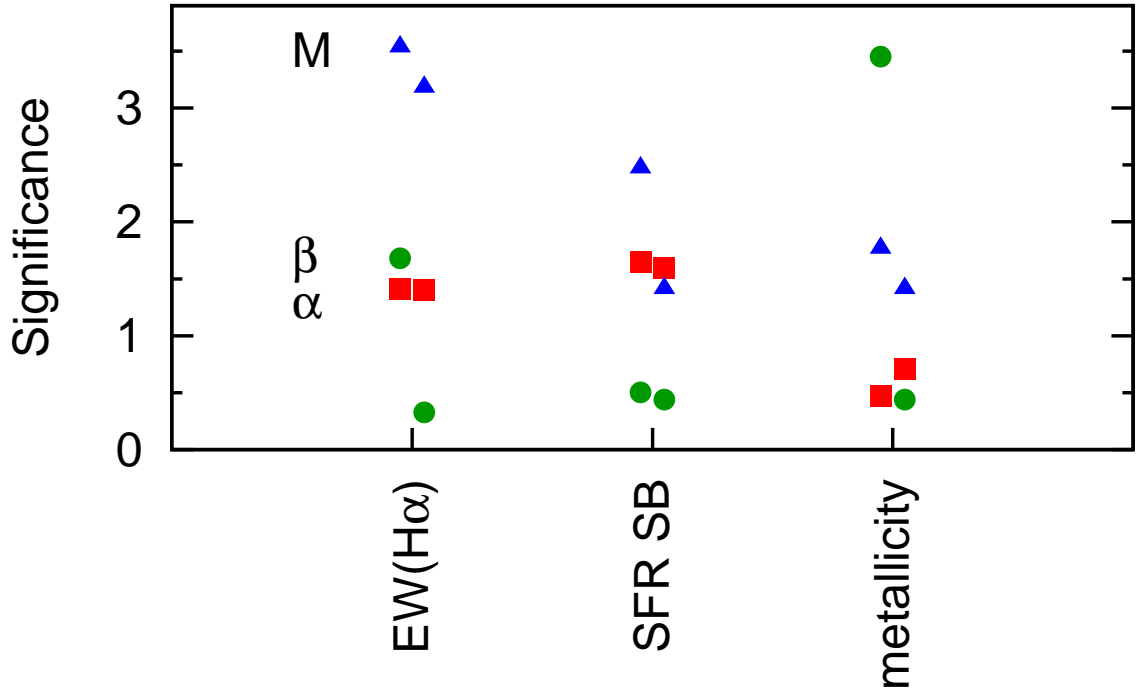}
 \caption{Significance levels in the differences of correction coefficients $\alpha$ (rectangles), $\beta$ (circles) and $M$ (triangles) between low/high EW {\ha}, SFR SD and metallicity. Left (right) side points of each entry are the values for the sample with (without) the color cut.
\label{alpbetf}}
\end{figure}

{\bf Progenitor model: } It is well documented that CO WDs in a binary system can 
increase their masses by accretion from their companion star \citep[e.g.][]{whe73,ibe84}. 
When their masses reach the Chandrasekhar limit, they are expected to explode as SNe Ia.
An explosion model by \citet{hac96} proposed that WDs below a metallicity 
cutoff can not explode as SNe~Ia because they cannot produce a wind.  
Based on this scenario, \citet{kob98} predicted a metallicity cutoff of 
[Fe/H] $= -1.1$.  This value corresponds to {\logoh} $=8.0$ by 
Equation \ref{feh} and to {\logten}([{\NII}]{\lam 6585}/{\ha}) $=-1.65$ by 
the calibration formula of \citet{pet04}.
As shown in Figure \ref{n2ha_o3hb}, there are no SN Ia hosts below the metallicity of $8.0$.
Since the MPA/JHA database itself has a very low fraction ($< 0.1$ \%) of 
such galaxies  (Figure \ref{n2ha_o3hb}), the absence of SN Ia hosts might 
not be meaningful, but, they might support the Hachisu model.

Metal-poor hosts below the metallicity cutoff were obtained in the high-z 
samples of \citet{how09} and \citet{sul10}.  However, the average 
metallicity drops by at most 0.1 dex \citep{rod08} between nearby and 
their largest redshift hosts. Thus, it is hard to believe that the 
existence of metal-poor hosts does not result from the usage of nearby 
empirical relations between stellar masses and metallicities
\citep{tre04,lia06}.  These hosts might imply a different channel to 
the SN Ia explosion from the Hachisu model.  Spectroscopic observations 
of those galaxies will be of great help to discuss their metallicity effects.

\section{Conclusions} \label{sum09d}
We have analyzed the multi-band lightcurves of 118 confirmed SNe Ia and 
the spectra of their host galaxies.  We derived the EW {\ha}, 
SFR SD, and gas-phase metallicity from the spectra and compared these 
with the lightcurve widths and colors of SNe Ia.  In addition, we 
compared host properties with the Hubble residuals corrected for 
lightcurve parameters to investigate uncorrected systematic effects in 
the magnitude standardization. We conclude the following:

(i) SNe Ia in hosts with a higher star formation rate, on average, 
have synthesized larger {\rni} mass and show wider lightcurves. The {\rni} 
mass dependence is consistent with a nucleosynthesis-based prediction. 

(ii) SNe Ia in metal-rich galaxies ({\logoh}$>9.0$) have become 0.13 
magnitude brighter (at the 1.8 $\sigma$ level) after lightcurve corrections, 
which corresponds to up to 6 \% uncertainty in the luminosity distance.

(iii) The coefficient of the color correction term in standardizing 
luminosity is larger for SNe Ia in metal-poor hosts or hosts with a large 
EW {\ha} (at the $\sim 2\sigma$ level).

\acknowledgements
Acknowledgements -- 
K.K. thanks the COE Program ``the Quantum Extreme Systems and Their Symmetries'' for fiscal 2007, the Global COE Program ``the Physical Sciences Frontier'' for fiscal 2008-2010, MEXT, Japan and the JASSO scholarship for fiscal 2007-2009. 

Funding for the SDSS and SDSS-II was provided by the Alfred P. Sloan Foundation, the Participating Institutions, the National Science Foundation, the U.S. Department of Energy, the National Aeronautics and Space Administration, the Japanese Monbukagakusho, the Max Planck Society, and the Higher Education Funding Council for England. \url{The SDSS Web site is http://www.sdss.org/}.

The SDSS is managed by the Astrophysical Research Consortium (ARC) for the Participating Institutions. The Participating Institutions are The University of Chicago, Fermilab, the Institute for Advanced Study, the Japan Participation Group, The Johns Hopkins University, Los Alamos National Laboratory, the Max-Planck-Institute for Astronomy (MPIA), the Max-Planck-Institute for Astrophysics (MPA), New Mexico State University, University of Pittsburgh, Princeton University, the United States Naval Observatory, and the University of Washington. This research has made use of the NASA/IPAC Extragalactic Database (NED) which is operated by the Jet Propulsion Laboratory, California Institute of Technology, under contract with the National Aeronautics and Space Administration. 

Facilities: \facility{SDSS}

\clearpage

\clearpage
\begin{deluxetable}{llrrrrrl} 
 \tabletypesize{\scriptsize} 
 \tablecaption{
 Summary of spectroscopic properties of the SDSS host galaxy\label{spec_sdss}} 
 \tablewidth{0pt} 
% \rotate 
 \tablehead{ 
 \colhead{ID$^a$} & \colhead{IAU name} & \colhead{EW ({\ha})} & \colhead{{\ebv}} & 
 \colhead{SFR [{\Msunyr}~kpc$^{-2}$]} &  \colhead{\logoh} & 
 \colhead{Calib$^b$} & \colhead{Env$^c$}
 }
 \startdata 
722   & 2005ed &   $1.99 \pm  0.17$ & $-0.00 \pm 0.26$ & $(9.53 \pm 1.04) \times 10^{-2}$ &  $9.045 \pm 0.042$ &  N2/O2a & AGN \\ 
739   & 2005ef &   $1.58 \pm  0.26$ &              --- & $(6.53 \pm 1.13) \times 10^{-2}$ &  $9.200 \pm 0.046$ &  N2/O2g & N/A \\ 
762   & 2005eg &   $9.92 \pm  2.12$ &  $0.49 \pm 0.39$ & $(1.96 \pm 0.57) \times 10^{-1}$ &  $9.087 \pm 0.081$ &  N2/O2a & AGN \\ 
774   & 2005ex &   $5.09 \pm  0.26$ &  $1.28 \pm 0.40$ & $(2.99 \pm 0.13) \times 10^{-1}$ &  $8.658 \pm 0.086$ &  N2/O2a & AGN \\ 
1032  & 2005ez &   $3.14 \pm  0.27$ & $-0.23 \pm 0.20$ & $(5.37 \pm 0.46) \times 10^{-2}$ &  $9.069 \pm 0.045$ &  N2/O2a & AGN \\ 
1112  & 2005fg &  $11.71 \pm  1.48$ &  $0.48 \pm 0.29$ & $(1.64 \pm 0.18) \times 10^{-1}$ &  $9.101 \pm 0.040$ &  N2/O2g & AGN \\ 
1371  & 2005fh &   $0.44 \pm  0.14$ & $-0.21 \pm 0.23$ & $(2.85 \pm 0.56) \times 10^{-2}$ &  $8.918 \pm 0.288$ &  N2/S2h & AGN \\ 
1580  & 2005fb &  $48.40 \pm  0.33$ &  $0.54 \pm 0.04$ & $(1.98 \pm 0.02) \times 10^{ 0}$ &  $9.085 \pm 0.003$ &  N2/O2a & AGN \\ 
2561  & 2005fv &  $12.12 \pm  0.53$ &  $0.75 \pm 0.04$ & $(1.54 \pm 0.05) \times 10^{-1}$ &  $8.836 \pm 0.023$ &  N2/O2a &  SF \\ 
2689  & 2005fa &   $1.57 \pm  0.44$ &              --- & $(7.65 \pm 1.88) \times 10^{-2}$ &                N/A &   caseX & N/A \\ 
2992  & 2005gp &  $13.19 \pm  0.66$ &  $0.49 \pm 0.25$ & $(8.43 \pm 0.77) \times 10^{-2}$ &  $8.992 \pm 0.034$ &  N2/O2a &  SF \\ 
3592  & 2005gb &  $18.72 \pm  0.30$ &  $0.64 \pm 0.05$ & $(3.79 \pm 0.06) \times 10^{-1}$ &  $9.048 \pm 0.018$ &  N2/O2a & AGN \\ 
3901  & 2005ho &  $59.82 \pm  0.61$ &  $0.22 \pm 0.02$ & $(6.89 \pm 0.06) \times 10^{-1}$ &  $8.649 \pm 0.012$ &  N2/O2a &  SF \\ 
5944  & 2005hc &   $1.63 \pm  0.19$ & $-0.16 \pm 0.21$ & $(4.80 \pm 0.54) \times 10^{-2}$ &  $8.789 \pm 0.123$ &  N2/S2h & AGN \\ 
5966  & 2005it &  $17.58 \pm  0.69$ &  $0.18 \pm 0.09$ & $(2.19 \pm 0.10) \times 10^{-1}$ &  $9.262 \pm 0.024$ &  N2/O2a &  SF \\ 
6057  & 2005if &  $34.25 \pm  0.32$ &  $0.19 \pm 0.03$ & $(4.26 \pm 0.02) \times 10^{-1}$ &  $8.977 \pm 0.014$ &  N2/O2a &  SF \\ 
6295  & 2005js &   $0.52 \pm  0.14$ &              --- & $(3.93 \pm 1.32) \times 10^{-2}$ &                N/A &   caseX & N/A \\ 
6406  & 2005ij &  $13.62 \pm  0.48$ &  $0.49 \pm 0.11$ & $(1.59 \pm 0.04) \times 10^{-1}$ &  $8.898 \pm 0.031$ &  N2/O2a &  SF \\ 
7876  & 2005ir &  $20.91 \pm  0.39$ &  $0.32 \pm 0.05$ & $(2.92 \pm 0.05) \times 10^{-1}$ &  $9.039 \pm 0.016$ &  N2/O2a &  SF \\ 
8151  & 2005hk &   $9.71 \pm  0.33$ &  $0.06 \pm 0.07$ & $(1.55 \pm 0.04) \times 10^{-1}$ &  $8.229 \pm 0.050$ &  N2/S2h & N/A \\ 
10028 & 2005kt &   $0.35 \pm  0.13$ &              --- & $(2.12 \pm 0.36) \times 10^{-2}$ &  $8.994 \pm 0.053$ &  N2/O2g & N/A \\ 
10096 & 2005lj &  $22.92 \pm  0.58$ &  $0.16 \pm 0.05$ & $(2.03 \pm 0.04) \times 10^{-1}$ &  $8.886 \pm 0.017$ &  N2/O2a &  SF \\ 
10434 & 2005lk &   $3.83 \pm  0.35$ &  $0.54 \pm 0.24$ & $(9.04 \pm 0.78) \times 10^{-2}$ &  $8.761 \pm 0.102$ &  N2/S2h & AGN \\ 
10805 & 2005ku &  $37.62 \pm  0.04$ &  $0.32 \pm 0.02$ & $(1.28 \pm 0.00) \times 10^{ 0}$ &  $8.919 \pm 0.003$ &  N2/O2a &  SF \\ 
12778 & 2006fs &  $20.26 \pm  0.33$ &  $0.40 \pm 0.06$ & $(4.79 \pm 0.08) \times 10^{-1}$ &  $8.953 \pm 0.018$ &  N2/O2a &  SF \\ 
12781 & 2006er &   $0.77 \pm  0.16$ & $-0.46 \pm 0.15$ & $(2.68 \pm 0.44) \times 10^{-2}$ &  $9.278 \pm 0.050$ &  N2/O2a &  SF \\ 
12843 & 2006fa &   $0.46 \pm  0.27$ & $-0.67 \pm 0.58$ & $(1.26 \pm 0.48) \times 10^{-2}$ &  $9.256 \pm 0.045$ &  N2/O2d &  SF \\ 
12856 & 2006fl &  $23.95 \pm  0.95$ &  $0.18 \pm 0.11$ & $(1.76 \pm 0.05) \times 10^{-1}$ &  $8.937 \pm 0.054$ &  N2/O2a &  SF \\ 
12874 & 2006fb &  $11.75 \pm  0.78$ &  $0.63 \pm 0.16$ & $(1.27 \pm 0.06) \times 10^{-1}$ &  $8.964 \pm 0.052$ &  N2/O2g & N/A \\ 
12897 & 2006eb &   $1.18 \pm  0.11$ &  $0.28 \pm 0.20$ & $(1.03 \pm 0.07) \times 10^{-1}$ &  $8.938 \pm 0.161$ &  N2/S2h & N/A \\ 
12950 & 2006fy &  $40.73 \pm  1.04$ &  $0.05 \pm 0.03$ & $(4.97 \pm 0.15) \times 10^{-1}$ &  $8.819 \pm 0.013$ &  N2/O2a &  SF \\ 
12971 & 2006ff &   $1.16 \pm  0.18$ &              --- & $(3.65 \pm 0.44) \times 10^{-2}$ &  $8.912 \pm 0.236$ &  N2/Hai & N/A \\ 
12979 & 2006gf &   $0.68 \pm  0.16$ &              --- & $(7.14 \pm 1.82) \times 10^{-2}$ &                N/A &   caseX & N/A \\ 
12983 & 2006gl &  $85.57 \pm  2.18$ &  $0.57 \pm 0.07$ & $(4.06 \pm 0.17) \times 10^{-1}$ &  $8.978 \pm 0.019$ &  N2/O2a &  SF \\ 
13070 & 2006fu &  $44.54 \pm  1.09$ &  $0.48 \pm 0.12$ & $(2.24 \pm 0.16) \times 10^{-1}$ &  $8.883 \pm 0.032$ &  N2/O2a &  SF \\ 
13072 & 2006fi & $243.83 \pm  6.44$ &  $0.19 \pm 0.00$ & $(4.19 \pm 0.00) \times 10^{ 0}$ &  $8.873 \pm 0.008$ &  N2/O2a &  SF \\ 
13099 & 2006gb &  $23.97 \pm  1.21$ &  $0.29 \pm 0.08$ & $(3.05 \pm 0.12) \times 10^{-1}$ &  $9.111 \pm 0.023$ &  N2/O2a &  SF \\ 
13135 & 2006fz &   $1.07 \pm  0.23$ &              --- & $(2.73 \pm 0.63) \times 10^{-2}$ &  $9.150 \pm 0.071$ &  N2/O2g & N/A \\ 
13254 & 2006gx &  $38.16 \pm  1.38$ &  $0.17 \pm 0.11$ & $(2.06 \pm 0.17) \times 10^{-1}$ &  $8.930 \pm 0.025$ &  N2/O2a &  SF \\ 
13354 & 2006hr &  $48.18 \pm  1.12$ &  $0.37 \pm 0.04$ & $(5.62 \pm 0.11) \times 10^{-1}$ &  $8.992 \pm 0.010$ &  N2/O2a &  SF \\ 
13511 & 2006hh &   $5.04 \pm  0.40$ &  $0.67 \pm 0.35$ & $(1.12 \pm 0.09) \times 10^{-1}$ &  $8.983 \pm 0.095$ &  N2/O2d & N/A \\ 
13610 & 2006hd &  $74.48 \pm  1.12$ &  $0.36 \pm 0.03$ & $(9.25 \pm 0.12) \times 10^{-1}$ &  $8.906 \pm 0.009$ &  N2/O2a &  SF \\ 
14279 & 2006hx &   $5.58 \pm  0.16$ &  $0.55 \pm 0.08$ & $(2.81 \pm 0.07) \times 10^{-1}$ &  $9.118 \pm 0.034$ &  N2/O2a & AGN \\ 
14284 & 2006ib &   $0.08 \pm  0.01$ &              --- &                              N/A &                N/A &   caseX & N/A \\ 
14318 & 2006py &   $3.81 \pm  0.27$ &  $0.46 \pm 0.23$ & $(6.76 \pm 0.39) \times 10^{-2}$ &  $8.878 \pm 0.038$ &  N2/O2a & AGN \\ 
14421 & 2006ia &   $1.55 \pm  0.56$ &  $0.09 \pm 0.44$ & $(4.30 \pm 0.95) \times 10^{-2}$ &  $9.149 \pm 0.030$ &  N2/O2d &  SF \\ 
14816 & 2006ja &   $2.81 \pm  0.18$ &  $0.49 \pm 0.22$ & $(5.75 \pm 0.37) \times 10^{-2}$ &  $8.973 \pm 0.030$ &  N2/O2a & AGN \\ 
15129 & 2006kq &  $20.75 \pm  0.53$ &  $0.40 \pm 0.13$ & $(1.74 \pm 0.14) \times 10^{-1}$ &  $8.995 \pm 0.034$ &  N2/O2a &  SF \\ 
15136 & 2006ju &  $37.77 \pm  0.36$ &  $0.39 \pm 0.03$ & $(9.34 \pm 0.16) \times 10^{-1}$ &  $9.134 \pm 0.003$ &  N2/O2a &  SF \\ 
15161 & 2006jw &   $8.93 \pm  1.04$ &  $0.30 \pm 0.22$ & $(8.30 \pm 0.75) \times 10^{-2}$ &  $9.080 \pm 0.098$ &  N2/O2a &  SF \\ 
15222 & 2006jz &   $6.41 \pm  0.64$ &              --- & $(1.78 \pm 0.17) \times 10^{-1}$ &  $8.617 \pm 0.193$ &  N2/S2h & N/A \\ 
15234 & 2006kd &  $15.35 \pm  0.61$ &  $0.26 \pm 0.10$ & $(1.55 \pm 0.05) \times 10^{-1}$ &  $9.089 \pm 0.033$ &  N2/O2a &  SF \\ 
15421 & 2006kw &  $40.93 \pm  1.21$ &  $0.08 \pm 0.05$ & $(2.42 \pm 0.05) \times 10^{-1}$ &  $8.865 \pm 0.017$ &  N2/O2a &  SF \\ 
15425 & 2006kx &   $2.20 \pm  0.57$ &              --- & $(2.14 \pm 0.64) \times 10^{-2}$ &  $9.084 \pm 0.032$ &  N2/O2g & N/A \\ 
15443 & 2006lb &  $32.02 \pm  0.78$ &  $0.26 \pm 0.07$ & $(2.67 \pm 0.08) \times 10^{-1}$ &  $8.930 \pm 0.023$ &  N2/O2a &  SF \\ 
15467 &    ---$^d$ &  $35.30 \pm  1.03$ &  $0.27 \pm 0.08$ & $(3.99 \pm 0.10) \times 10^{-1}$ &  $8.984 \pm 0.019$ &  N2/O2a &  SF \\ 
15648 & 2006ni &   $1.23 \pm  0.31$ & $-0.53 \pm 0.62$ & $(2.42 \pm 1.07) \times 10^{-2}$ &  $9.208 \pm 0.064$ &  N2/O2g & N/A \\ 
15734 & 2006ng &  $44.91 \pm  0.37$ &  $0.16 \pm 0.02$ & $(1.17 \pm 0.02) \times 10^{ 0}$ &  $8.641 \pm 0.017$ &  N2/O2a &  SF \\ 
16069 & 2006nd &  $33.51 \pm  0.48$ &  $0.61 \pm 0.04$ & $(7.37 \pm 0.10) \times 10^{-1}$ &  $9.037 \pm 0.019$ &  N2/O2a &  SF \\ 
16099 & 2006nn &  $-1.85 \pm  1.27$ &              --- &                              N/A &  $9.099 \pm 0.040$ &  N2/O2g & N/A \\ 
16211 & 2006nm &   $0.40 \pm  1.08$ &              --- & $(8.03 \pm 94.20) \times 10^{-4}$ &                N/A &     N/A & N/A \\ 
16215 & 2006ne &   $7.56 \pm  0.29$ &  $0.36 \pm 0.09$ & $(1.19 \pm 0.04) \times 10^{-1}$ &  $9.142 \pm 0.048$ &  N2/O2a &  SF \\ 
16259 & 2006ol &   $1.66 \pm  0.22$ &  $0.03 \pm 0.38$ & $(5.62 \pm 0.97) \times 10^{-2}$ &  $9.052 \pm 0.051$ &  N2/O2a &  SF \\ 
16280 & 2006nz &   $0.58 \pm  0.13$ &              --- & $(5.69 \pm 1.29) \times 10^{-2}$ &                N/A &   caseX & N/A \\ 
16314 & 2006oa &  $29.52 \pm  1.50$ &  $0.24 \pm 0.10$ & $(4.78 \pm 0.17) \times 10^{-2}$ &  $8.730 \pm 0.060$ &  N2/O2a &  SF \\ 
16333 & 2006on &   $0.28 \pm  0.19$ & $-0.20 \pm 0.31$ & $(2.60 \pm 0.28) \times 10^{-2}$ &  $9.124 \pm 0.041$ &  N2/O2a &  SF \\ 
16392 & 2006ob &   $1.04 \pm  0.12$ &  $0.20 \pm 0.30$ & $(1.60 \pm 0.19) \times 10^{-1}$ &  $9.092 \pm 0.058$ &  N2/O2a & AGN \\ 
16482 & 2006pm &   $0.55 \pm  0.25$ &              --- & $(1.69 \pm 0.65) \times 10^{-2}$ &                N/A &   caseX & N/A \\ 
16644 & 2006pt &  $19.69 \pm  2.14$ &  $0.53 \pm 0.05$ & $(2.36 \pm 0.02) \times 10^{-1}$ &  $9.036 \pm 0.021$ &  N2/O2a &  SF \\ 
16692 & 2006op &   $5.23 \pm  0.35$ &  $0.39 \pm 0.20$ & $(5.36 \pm 0.24) \times 10^{-2}$ &  $8.873 \pm 0.065$ &  N2/O2a & N/A \\ 
16789 & 2006pz &   $0.37 \pm  0.40$ &              --- & $(8.42 \pm 21.00) \times 10^{-3}$ &                N/A &     N/A & N/A \\ 
17117 & 2006qm & $104.18 \pm  1.25$ &  $0.20 \pm 0.01$ & $(1.19 \pm 0.01) \times 10^{ 0}$ &  $8.881 \pm 0.007$ &  N2/O2a &  SF \\ 
17134 &    ---$^d$ &  $15.33 \pm  0.49$ &  $0.42 \pm 0.04$ & $(0.00 \pm 0.00) \times 10^{ 0}$ &  $8.987 \pm 0.018$ &  N2/O2a &  SF \\ 
17135 & 2006rz &  $27.24 \pm  0.43$ &  $0.04 \pm 0.01$ & $(0.00 \pm 0.00) \times 10^{ 0}$ &  $8.563 \pm 0.020$ &  N2/O2a & N/A \\ 
17171 & 2007id &   $0.52 \pm  0.38$ &              --- & $(1.68 \pm 0.57) \times 10^{-2}$ &                N/A &  N2/Hai & N/A \\ 
17176 & 2007ie &  $49.41 \pm  0.11$ &  $0.11 \pm 0.04$ & $(2.33 \pm 0.01) \times 10^{-1}$ &  $8.438 \pm 0.046$ &  N2/O2a &  SF \\ 
17186 & 2007hx &   $9.77 \pm  0.32$ &  $0.58 \pm 0.14$ & $(1.15 \pm 0.04) \times 10^{-1}$ &  $8.934 \pm 0.038$ &  N2/O2a & AGN \\ 
17215 & 2007hy &   $1.38 \pm  0.01$ &              --- & $(2.51 \pm 0.73) \times 10^{-2}$ &  $9.091 \pm 0.038$ &  N2/O2g & N/A \\ 
17280 & 2007ia &  $20.88 \pm  0.26$ &  $0.39 \pm 0.02$ & $(6.10 \pm 0.02) \times 10^{-1}$ &  $9.122 \pm 0.029$ &  N2/O2a & AGN \\ 
17332 & 2007jk &   $8.63 \pm  0.91$ &  $0.36 \pm 0.29$ & $(5.87 \pm 0.54) \times 10^{-2}$ &  $7.773 \pm 0.321$ &  N2/S2h & AGN \\ 
17340 & 2007kl &   $4.71 \pm  0.38$ &              --- & $(6.83 \pm 0.62) \times 10^{-2}$ &  $9.093 \pm 0.063$ &  N2/O2g & N/A \\ 
17366 & 2007hz &  $12.61 \pm  0.38$ &  $0.37 \pm 0.09$ & $(2.56 \pm 0.07) \times 10^{-1}$ &  $9.109 \pm 0.048$ &  N2/O2g & N/A \\ 
17497 & 2007jt &  $25.44 \pm  0.80$ &  $0.40 \pm 0.08$ & $(2.41 \pm 0.06) \times 10^{-1}$ &  $9.011 \pm 0.023$ &  N2/O2g &  SF \\ 
17500 & 2007lf &   $1.87 \pm  0.12$ & $-0.20 \pm 0.12$ & $(2.15 \pm 0.15) \times 10^{-1}$ &  $9.130 \pm 0.018$ &  N2/O2a & AGN \\ 
17784 & 2007jg &  $18.48 \pm  0.08$ &  $0.17 \pm 0.06$ & $(7.11 \pm 0.02) \times 10^{-2}$ &  $8.768 \pm 0.062$ &  N2/O2a & N/A \\ 
17880 & 2007jd &  $17.69 \pm  0.25$ &  $0.27 \pm 0.06$ & $(2.57 \pm 0.02) \times 10^{-1}$ &  $9.117 \pm 0.065$ &  N2/O2a &  SF \\ 
17886 & 2007jh &   $0.29 \pm  0.01$ &              --- & $(3.28 \pm 1.68) \times 10^{-2}$ &                N/A &   caseX & N/A \\ 
18030 & 2007kq &  $70.34 \pm  2.03$ &  $0.25 \pm 0.05$ & $(3.20 \pm 0.06) \times 10^{-1}$ &  $8.556 \pm 0.052$ &  N2/O2a &  SF \\ 
18298 & 2007li &   $1.12 \pm  0.26$ &              --- & $(2.07 \pm 0.33) \times 10^{-2}$ &  $9.178 \pm 0.073$ &  N2/O2g & N/A \\ 
18612 & 2007lc &  $11.05 \pm  0.13$ &  $0.58 \pm 0.05$ & $(2.79 \pm 0.09) \times 10^{-1}$ &  $9.044 \pm 0.009$ &  N2/O2a & AGN \\ 
18643 & 2007lv &   $0.39 \pm  0.20$ &              --- & $(4.05 \pm 1.35) \times 10^{-2}$ &                N/A &   caseX & N/A \\ 
18697 & 2007ma &  $27.28 \pm  0.53$ &  $0.54 \pm 0.06$ & $(3.28 \pm 0.06) \times 10^{-1}$ &  $9.036 \pm 0.026$ &  N2/O2a &  SF \\ 
18721 & 2007mu &   $0.00 \pm -1.00$ & $-0.05 \pm 0.18$ & $(1.05 \pm 0.16) \times 10^{-1}$ &                N/A &   caseX & N/A \\ 
18751 & 2007ly &   $0.11 \pm  0.16$ &              --- & $(2.31 \pm 1.84) \times 10^{-3}$ &                N/A &     N/A & N/A \\ 
18809 & 2007mi &   $0.36 \pm  0.29$ &              --- & $(1.79 \pm 2.05) \times 10^{-2}$ &                N/A &     N/A & N/A \\ 
18835 & 2007mj &   $0.92 \pm  0.32$ &              --- & $(2.63 \pm 1.06) \times 10^{-2}$ &  $9.116 \pm 0.064$ &  N2/O2g & N/A \\ 
18855 & 2007mh &  $16.68 \pm  0.88$ &  $0.18 \pm 0.17$ & $(1.33 \pm 0.06) \times 10^{-1}$ &  $9.128 \pm 0.041$ &  N2/O2a &  SF \\ 
18890 & 2007mm &   $0.12 \pm  0.22$ & $-1.31 \pm 0.47$ & $(7.46 \pm 2.83) \times 10^{-3}$ &                N/A &  N2/Hai & N/A \\ 
18903 & 2007lr &  $11.85 \pm  0.47$ &  $0.64 \pm 0.11$ & $(2.23 \pm 0.06) \times 10^{-1}$ &  $9.108 \pm 0.043$ &  N2/O2a &  SF \\ 
19155 & 2007mn &  $10.12 \pm  0.04$ &  $0.46 \pm 0.05$ & $(1.90 \pm 0.03) \times 10^{-1}$ &  $8.933 \pm 0.027$ &  N2/O2a &  SF \\ 
19353 & 2007nj &  $16.18 \pm  0.51$ &  $0.59 \pm 0.11$ & $(1.83 \pm 0.05) \times 10^{-1}$ &  $8.985 \pm 0.034$ &  N2/O2g & N/A \\ 
19616 & 2007ok &  $44.96 \pm  0.27$ &  $0.61 \pm 0.01$ & $(1.08 \pm 0.01) \times 10^{ 0}$ &  $9.061 \pm 0.004$ &  N2/O2a &  SF \\ 
19626 & 2007ou &  $39.39 \pm  0.67$ &  $0.45 \pm 0.02$ & $(4.96 \pm 0.08) \times 10^{-1}$ &  $8.961 \pm 0.007$ &  N2/O2a &  SF \\ 
19794 & 2007oz &  $-4.42 \pm  1.09$ &              --- &                              N/A &                N/A &     N/A & N/A \\ 
19969 & 2007pt &  $54.19 \pm  0.74$ &  $0.33 \pm 0.03$ & $(7.38 \pm 0.09) \times 10^{-1}$ &  $9.044 \pm 0.009$ &  N2/O2a &  SF \\ 
20064 & 2007om &   $6.43 \pm  0.40$ &  $1.15 \pm 0.37$ & $(1.51 \pm 0.08) \times 10^{-1}$ &  $8.763 \pm 0.088$ &  N2/O2a & AGN \\ 
20208 & 2007qd &  $55.31 \pm  2.46$ &  $0.18 \pm 0.07$ & $(5.67 \pm 0.26) \times 10^{-1}$ &  $8.817 \pm 0.028$ &  N2/O2a &  SF \\ 
20420 & 2007qw & $129.65 \pm  1.76$ &  $0.08 \pm 0.03$ & $(5.23 \pm 0.06) \times 10^{-1}$ &  $8.424 \pm 0.039$ &  N2/O2a &  SF \\ 
20528 & 2007qr &  $18.18 \pm  0.51$ &  $0.59 \pm 0.11$ & $(3.66 \pm 0.01) \times 10^{-1}$ &  $9.042 \pm 0.008$ &  N2/O2a &  SF \\ 
20625 & 2007px &  $14.01 \pm  0.56$ &  $0.54 \pm 0.12$ & $(1.82 \pm 0.06) \times 10^{-1}$ &  $8.989 \pm 0.045$ &  N2/O2a & N/A \\ 
20718 & 2007rj &   $7.37 \pm  0.25$ &  $0.42 \pm 0.13$ & $(1.63 \pm 0.05) \times 10^{-1}$ &  $9.066 \pm 0.025$ &  N2/O2a & AGN \\ 
20889 & 2007py &   $1.16 \pm  0.69$ &              --- & $(1.28 \pm 0.84) \times 10^{-2}$ &                N/A &     N/A & N/A \\ 
21034 & 2007qa &  $38.00 \pm  0.60$ &  $0.40 \pm 0.04$ & $(5.79 \pm 0.09) \times 10^{-1}$ &  $9.054 \pm 0.009$ &  N2/O2a & AGN \\ 
21502 & 2007ra &  $11.32 \pm  0.20$ &  $0.58 \pm 0.07$ & $(4.90 \pm 0.09) \times 10^{-1}$ &  $8.876 \pm 0.015$ &  N2/O2a & AGN \\ 
21510 & 2007sh &  $14.19 \pm  0.52$ &  $0.30 \pm 0.09$ & $(1.70 \pm 0.05) \times 10^{-1}$ &  $9.168 \pm 0.035$ &  N2/O2a &  SF \\ 
21669 & 2007rs &   $1.91 \pm  0.20$ &              --- & $(5.72 \pm 0.69) \times 10^{-2}$ &  $9.171 \pm 0.035$ &  N2/O2g & N/A \\ 
21766 & 2007rc &  $31.66 \pm  0.99$ &  $0.37 \pm 0.05$ & $(1.03 \pm 0.05) \times 10^{ 0}$ &  $9.093 \pm 0.005$ &  N2/O2a &  SF \\ 
22075 & 2007si &   $0.19 \pm  0.19$ &              --- & $(1.01 \pm 0.87) \times 10^{-2}$ &                N/A &     N/A & N/A \\ 
 \enddata 
 \tablecomments{"N/A" is tagged for non detection and "---" for low S/N cases.} 
 \tablenotetext{a}{We attached the same ID for the host galaxies as the SNe Ia} 
% \tablenotetext{b}{These values are not corrected for {\ebv}.}
 \tablenotetext{b}{The scheme to calibrate the gas phase metallicity. Detail is written in the text.}
 \tablenotetext{c}{A type of host galaxies. SF stands for the star forming galaxy and AGN for the galaxies with AGN activities.}
 \tablenotetext{d}{No IAU names have been attached but identified as SNe Ia.}
\end{deluxetable} 

\begin{deluxetable}{lrrrrr}
\tablecaption{Standard parameters \label{alpbet}}
\tablewidth{0pt}
\tablehead{
 \colhead{Datasets} & \colhead{$N$} & \colhead{$\alpha$} & 
 \colhead{$\beta$} & \colhead{$M$} & \colhead{rms}
}
\startdata
Sample without color cut \\
\tableline
 EW {\ha} & 81 & 0.08 (0.02) & 3.00 (0.09) & -19.11 (0.01) & 0.187\\
 - low ($<12 {\AA}$) & 40 & 0.13 (0.03) & 2.91 (0.11) & -19.18 (0.02) & 0.192\\
 - high ($>12 {\AA}$)& 41 & 0.07 (0.03) & 3.25 (0.17) & -19.08 (0.02) & 0.162\\
 SFR surface density & 83 & 0.09 (0.02) & 3.06 (0.09) & -19.11 (0.01) & 0.188\\
 - low ($<1.1 \times 10^{-2}${\Msunyr}~kpc$^{-2}$) & 41 & 0.13 (0.03) & 3.03 (0.13) & -19.16 (0.02) & 0.198\\
 - high ($>1.1 \times 10^{-2}${\Msunyr}~kpc$^{-2}$)& 42 & 0.06 (0.03) & 2.93 (0.15) & -19.09 (0.02) & 0.167\\
 metallicity & 72 & 0.07 (0.02) & 2.74 (0.09) & -19.09 (0.01) & 0.177\\
 - low  ({\logoh}$<9.0$) & 36 & 0.07 (0.03) & 3.21 (0.16) & -19.07 (0.02) & 0.189\\
 - high ({\logoh}$>9.0$) & 36 & 0.09 (0.03) & 2.54 (0.11) & -19.12 (0.02) & 0.138\\
\tableline\tableline
Sample with {\csalt}$<0.3$ \\
\tableline
 EW {\ha} & 73 & 0.09 (0.02) & 2.53 (0.16) & -19.10 (0.01) & 0.163\\
 - low ($<13 {\AA}$) & 36 & 0.14 (0.04) & 2.38 (0.27) & -19.17 (0.02) & 0.168\\
 - high($>13 {\AA}$) & 37 & 0.07 (0.03) & 2.49 (0.20) & -19.08 (0.02) & 0.141\\
 SFR surface density & 73 & 0.10 (0.02) & 2.53 (0.16) & -19.10 (0.01) & 0.160\\
 - low ($<1.77 \times 10^{-1}${\Msunyr}~kpc$^{-2}$) & 36 & 0.14 (0.03) & 2.53 (0.23) & -19.14 (0.02) & 0.176\\
 - high($>1.77 \times 10^{-1}${\Msunyr}~kpc$^{-2}$) & 37 & 0.06 (0.04) & 2.39 (0.22) & -19.10 (0.02) & 0.129\\
 metallicity & 67 & 0.09 (0.02) & 2.35 (0.15) & -19.09 (0.01) & 0.153\\
 - low	({\logoh}$<9.0$) & 33 & 0.08 (0.03) & 2.41 (0.21) & -19.08 (0.02) & 0.165\\
 - high ({\logoh}$>9.0$) & 34 & 0.11 (0.03) & 2.28 (0.21) & -19.12 (0.02) & 0.135\\
\enddata
\tablecomments{Standard parameters and a scatter of the Hubble residual are calculated to each sample for dependences on host characteristics: metallicity, SFR surface density and EW {\ha}. Then each sample is halved by their host characteristics to examine a possible anomaly between the low and high subsamples. Uncertainties are listed within parenthesis.}
\end{deluxetable}

\clearpage

\end{document}